\documentclass[conference]{IEEEtran}
\IEEEoverridecommandlockouts

\usepackage{cite}
\usepackage{amsmath,amssymb,amsfonts}
\usepackage{algorithmic}
\usepackage{graphicx}
\usepackage{textcomp}
\usepackage[table,xcdraw]{xcolor}
\usepackage[hyphens]{url}

\usepackage{pifont}
\usepackage{mdframed}
\usepackage{makecell}
\usepackage[ruled,linesnumbered,vlined]{algorithm2e} 
\usepackage{siunitx}  
\usepackage[most]{tcolorbox} 
\usepackage{booktabs}
\usepackage{tikz}
\newcommand{\circlednumber}[3][white]{%
  \tikz[baseline=(char.base)]{
    \node[
    shape=circle, 
    fill=#2, 
    text=#1, 
    draw=#1,
    inner sep=0.6pt] (char) {#3};
  }%
}

\definecolor{myred}{HTML}{AE6864} %
\definecolor{myblue}{HTML}{0F52BA} %
\definecolor{mygreen}{HTML}{42A246} %
\usepackage[]{hyperref}
\usepackage[english]{babel}
\usepackage{hyphenat}

\usepackage{tabularx}

\def\BibTeX{{\rm B\kern-.05em{\sc i\kern-.025em b}\kern-.08em
    T\kern-.1667em\lower.7ex\hbox{E}\kern-.125emX}}
\begin{document}

\pdfpagewidth=8.5in
\pdfpageheight=11in

\title{Patterns behind Chaos: Forecasting Data Movement for Efficient Large-Scale MoE LLM Inference}

\author{
\IEEEauthorblockN{Zhongkai Yu}
\IEEEauthorblockA{\textit{UCSD}\\
La Jolla, USA\\
zhy055@ucsd.edu}
\and
\IEEEauthorblockN{Yue Guan}
\IEEEauthorblockA{\textit{UCSD}\\
La Jolla, USA\\
y9guan@ucsd.edu}
\and
\IEEEauthorblockN{Zihao Yu}
\IEEEauthorblockA{\textit{Indiana University Bloomington}\\
Bloomington, USA\\
yuzih@iu.edu}
\and
\IEEEauthorblockN{Chenyang Zhou}
\IEEEauthorblockA{\textit{Columbia University}\\
New York, USA\\
cz2791@columbia.edu}
\and
\IEEEauthorblockN{Zhengding Hu}
\IEEEauthorblockA{\textit{UCSD}\\
La Jolla, USA\\
zhh068@ucsd.edu}
\and
\IEEEauthorblockN{Shuyi Pei}
\IEEEauthorblockA{\textit{Samsung Semiconductor}\\
San Jose, USA\\
shuyi.pei@samsung.com}
\and
\IEEEauthorblockN{Yangwook Kang}
\IEEEauthorblockA{\textit{Samsung Semiconductor}\\
San Jose, USA\\
yangwook.k@samsung.com}
\and
\IEEEauthorblockN{Yufei Ding}
\IEEEauthorblockA{\textit{UCSD}\\
La Jolla, USA\\
yufeiding@ucsd.edu}
\and
\IEEEauthorblockN{Po-An Tsai}
\IEEEauthorblockA{\textit{NVIDIA}\\
Santa Clara, USA\\
poant@nvidia.com}
}

\maketitle
\footnotetext{Accepted to ISCA 2026. This is the authors' preprint version.}

\begin{abstract}
Large-scale Mixture of Experts (MoE) Large Language Models (LLMs) have recently become the frontier open-weight models, achieving remarkable model capability similar to proprietary ones.
But their random expert selection mechanism introduces significant data movement overhead that becomes the dominant bottleneck in multi-unit LLM serving systems. 

To understand the patterns underlying this data movement, we conduct comprehensive data-movement-centric profiling across four state-of-the-art large-scale MoE models released in 2025  (200B-1000B) using over 24,000 requests spanning diverse workloads. 
We perform systematic analysis from both temporal and spatial perspectives and distill six key insights to guide the design of diverse serving systems.
We verify these insights on both future wafer-scale GPU
architectures and existing GPU systems. 
On wafer-scale GPUs, lightweight architectural modifications guided by
our insights yield a 6.6x average speedup across four
200B--1000B models. 
On existing GPU systems, our insights drive the
design of a prefill-aware expert placement algorithm that achieves up to 1.25x speedup on MoE computation.
Our work presents the first comprehensive data-centric analysis of
large-scale MoE models together with a concrete design study applying the
learned lessons. Our profiling traces are publicly available at \href{https://huggingface.co/datasets/core12345/MoE_expert_selection_trace}{\textcolor{blue}{\nolinkurl{https://huggingface.co/datasets/core12345/MoE_expert_selection_trace}}}.

\end{abstract}



\section{Introduction}

\begin{figure}[t!]
    \centering
    \includegraphics[width=0.49\textwidth]{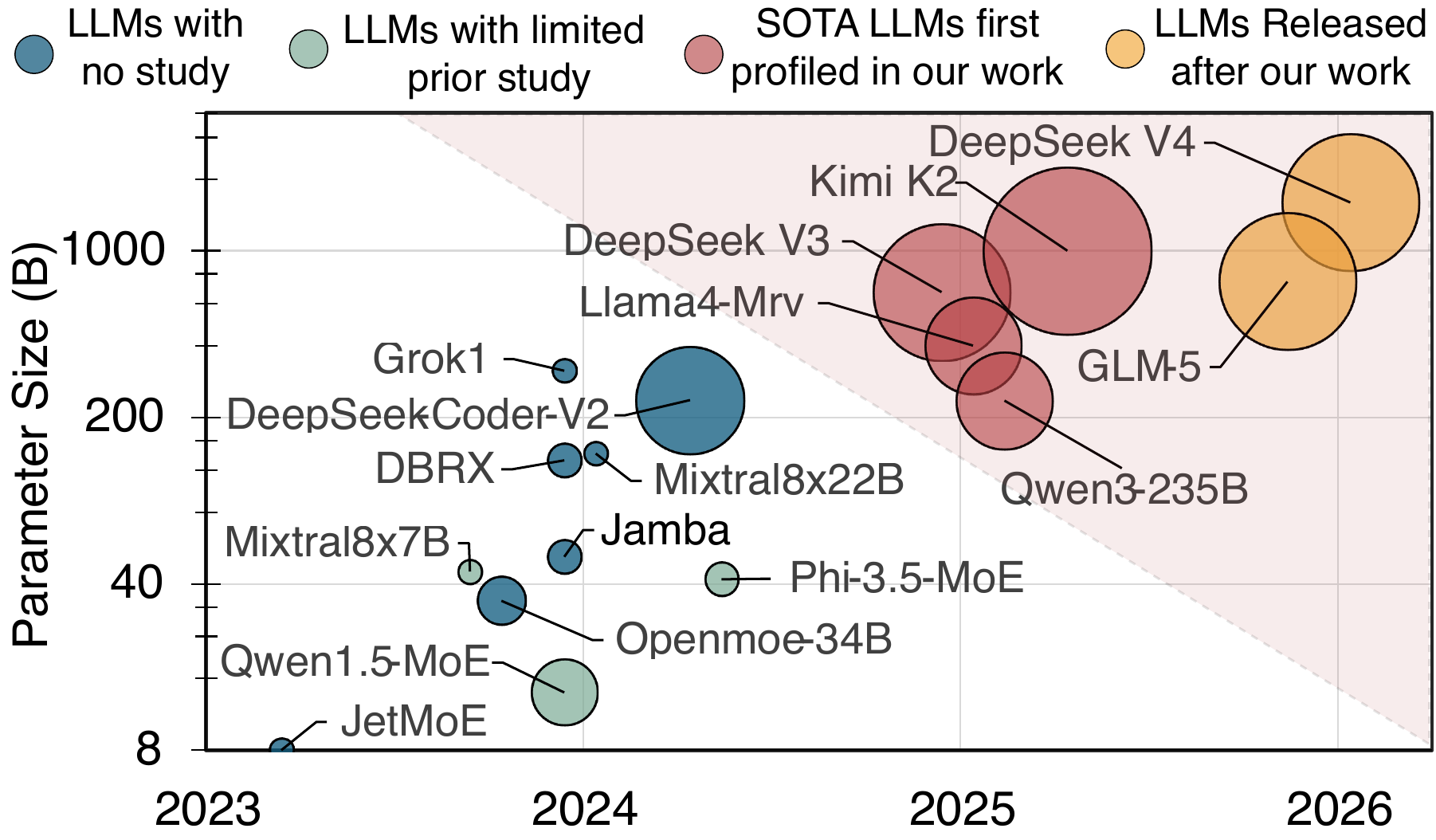}
    \caption{MoE LLM models sizes and release dates. Bubble size indicates the number of experts in each layer. Prior studies ~\cite{zhu2025megascale-infer, tairin2025emoe, skliar2024mixture, chitty2025lexi} provide limited analysis of smaller models from narrow perspectives, while our work presents the first comprehensive analysis of multiple unstudied SOTA models.}
    \label{fig:1_moe_survey}
\vspace{-12pt}
\end{figure}

Large Language Models (LLMs) have demonstrated remarkable capabilities across diverse domains, including programming assistance~\cite{llm_programming1, llm_programming2}, translation~\cite{gpt_llm_trans1, lu2024llm_trans2}, and chatbots~\cite{llm_chatbot1, llm_chatbot2}. 
Since the beginning of 2025, large-scale Mixture of Experts (MoE) LLMs (200B+ model with 100+ experts) have become the leading models for frontier LLMs~\cite{chiang2024chatbot} and the most widely used open weight models.
Unlike dense LLMs that activate all model weights uniformly, MoE models dynamically route each token to only a subset of experts, introducing substantial data movement overhead. Such overhead already exceeds 50\% of execution time for small models (e.g., Mixtral 8x7B) on modest systems (2--4 GPUs), and it exacerbates further with larger models such as DeepSeek V3 with 32$\times$ experts and 15$\times$ parameters deployed on multi-node systems (32+ GPUs)~\cite{go2025moetuner, zhang2025comet}. Moreover, this scaling trend is accelerating: recent releases such as DeepSeek V4~\cite{deepseekai2026deepseekv4} and GLM-5~\cite{glm5team2026glm5} continue to push the frontier, making the associated data movement patterns ever more critical. Yet as shown in~\autoref{fig:1_moe_survey}, no prior work has systematically investigated these patterns at scale. Earlier studies~\cite{fang2025klotski,tairin2025emoe,yao2024exploiting} confined themselves to profiling one or two small MoEs on limited hardware, reporting surface-level observations without system-level insights. As parameter sizes and expert counts surge, new data movement patterns have emerged but remain unexplored, leaving significant optimization opportunities on the table. \textbf{A comprehensive characterization of data movement in SOTA MoE models therefore presents a fruitful opportunity for better efficiency.}

If data movement in MoE models were fully unpredictable, it would present significant challenges for deployments on multi-unit systems.
\textbf{From a temporal perspective}, the explosive growth in expert combinations would make it impossible to prefetch, cache, or replicate experts in advance. 
For example, large-scale MoE models like DeepSeek V3 have $C_{256}^8 = \num{4426165368}$ combinations in expert selection.
When served with host memory-offload systems, such unpredictability would result in data movement like expert migrations between GPU and host, incurring substantial overhead, as inter-unit communication becomes the primary bottleneck.
\textbf{From a spatial perspective}, if expert selection were truly random, it would lead to severe workload imbalance across units. When queries from diverse tasks are served concurrently, the number of queries assigned to each expert would vary dramatically, creating significant workload disparities. Consequently, most units would remain idle and wait for heavily loaded units to finish, resulting in poor hardware resource utilization.

Fortunately, as we later show in the paper, \textbf{MoE expert selections indeed have predictability that designers can exploit to reduce data movement.} To uncover the inherent patterns in MoE models, we conduct a comprehensive data-movement-centric profiling of four state-of-the-art MoE models ranging from 235B to 1000B parameters released in 2025. 
As highlighted in~\autoref{fig:1_moe_survey}, we profile DeepSeek V3~\cite{liu2024deepseek}, Llama4 Maverick~\cite{llama4}, Qwen3-235B~\cite{yang2025qwen3}, and Kimi K2~\cite{team2025kimi} across 24,000 requests involving varied tasks, topics, and languages, which consumes $>$2000 GPU hours in total.
We then collect the expert selection trace of all layers and tokens in each request to create an expert selection database of over 150 GB JSON files. 
From these extensive traces, we conduct a comprehensive analysis to uncover data movement patterns from both temporal and spatial perspectives, making our findings \emph{system-agnostic} and applicable to various serving architectures at any scale. 
We then distill six key insights that serve as a solid foundation to understand MoE data movement and directly inform future MoE LLM serving system design, addressing critical questions that have remained unanswered in the field, such as: \emph{Is there a correlation between previously selected experts and those selected later? Are there discernible rules underlying the observed expert selection skewness? Do different tasks tend to activate different experts?} Our work represents the first systematic effort to characterize data movement patterns at the scale of up-to 1000B model across a wide range of tasks, providing actionable insights that can guide the design of next-generation MoE serving systems.


To demonstrate the broad applicability of our insights, we present case
studies on both future and existing GPU systems. On the architecture
side, we observe that modern GPUs have already adopted multi-chiplet
designs due to single-die size limitations~\cite{mi300, cpelide,
blackwell} and are evolving toward wafer-scale integration enabled by
emerging on-wafer packaging technologies~\cite{he2025waferlm,
xu2025wsc}. Targeting this trend, we develop a two-level
data-placement-aware command processor and a hardware-managed HBM scheme
that jointly balance workload across dies and reduce inter-die
communication, achieving an average 6.6$\times$ speedup in MoE serving
throughput on wafer-scale GPUs. On existing multi-GPU systems, we observe
that prefill-stage expert selections can effectively predict decode-stage
behavior. Building on this observation, we propose prefill-aware expert
placement algorithms to reduce decode workload imbalance,
and achieve up to 1.25$\times$ speedup.
Our main contributions can be summarized as follows:
\begin{itemize}
    \item We propose a comprehensive and systematic data-movement-centric profiling across four latest, large-scale MoE models released in 2025 between 235B and 1000B to uncover the data movement patterns from both temporal and spatial perspectives.
    \item We distill six key insights for designing efficient MoE serving systems based on our profiling and analysis, providing actionable guidance that can inspire future research in MoE serving systems.
    \item Leveraging these insights, we conduct case studies on both future and existing GPU systems. On future wafer-scale GPUs, we improve MoE throughput by 6.6$\times$ with minor hardware modifications. On existing multi-GPU systems, we achieve up to 1.25$\times$ speedup on an 8$\times$H100.
    \item We collect over 70,000 expert selection traces across multiple models and datasets, totaling over 150 GB in JSON format, and have open-sourced all traces with our multi-chiplet simulator to facilitate future research.
\end{itemize}



\begin{figure}[t]
    \centering
    \includegraphics[width=0.49\textwidth]{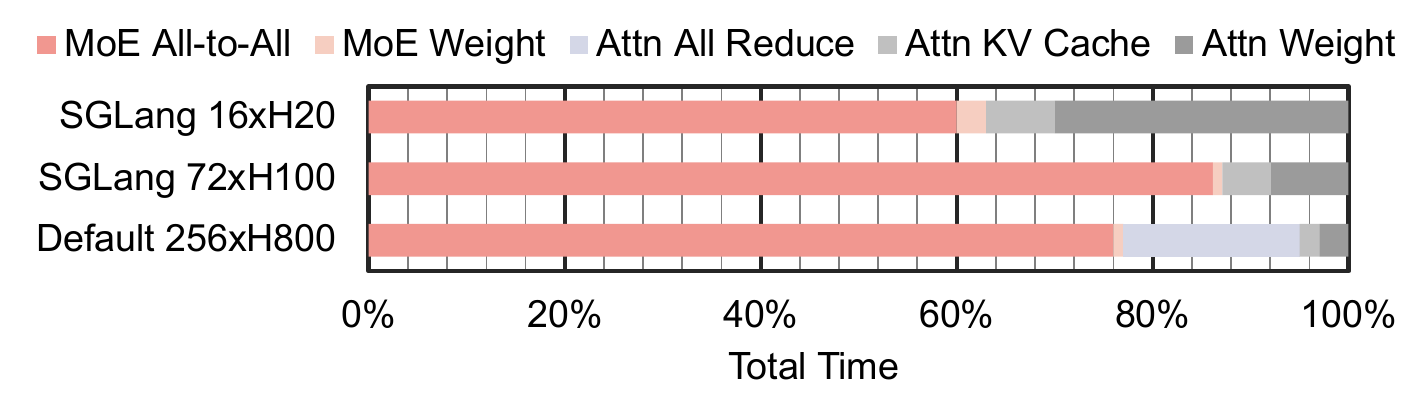}
    \caption{Latency breakdown for different data movement in DeepSeekV3 (4K sequence), modeled after various serving configurations~\cite{ds_sglang_16h20, sglang-deepseek-blogpost, liu2024deepseek}.
    }
    \label{fig:1-breakdown}
\vspace{-12pt}
\end{figure}

\begin{figure*}[t!]
    \centering
    \includegraphics[width=0.95\textwidth]{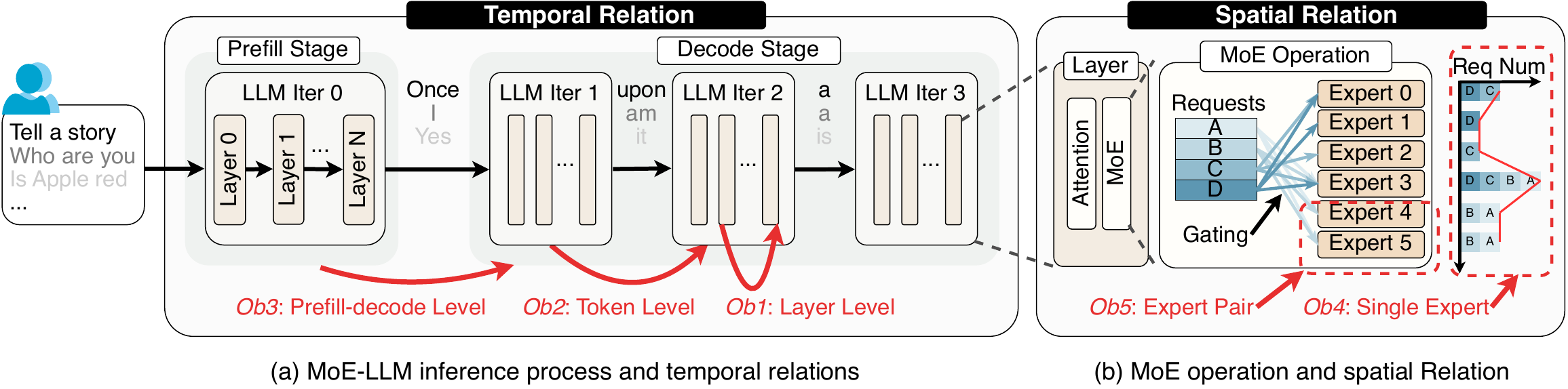}
    \caption{Inference process of MoE LLMs and the categorization method for our proposed data-centric profiling approach.}
    \label{fig:2_intro}
\end{figure*}

\section{Background}
\subsection{LLM and MoE Model Architecture}
Most state-of-the-art LLMs adopt a decoder-only transformer architecture that follows a token-by-token auto-regressive workflow~\cite{attention}.
As shown in ~\autoref{fig:2_intro}(a), after users input queries, the serving process is divided into two stages: the prefill stage and the decode stage.
During the prefill stage, all input tokens are processed simultaneously to generate the first output token.
The decode stage follows immediately, where tokens are generated sequentially. The generated token from each iteration is appended to the input sequence to produce the next token in the following iteration.

The Mixture of Experts (MoE) mechanism is a state-of-the-art approach to improve LLM performance and has become prevalent among current frontier LLMs~\cite{moe-survey}. As shown in ~\autoref{fig:2_intro}(b), MoE-based LLMs replace the feed-forward network (FFN) layers in traditional LLMs with MoE layers. In each layer, multiple experts are deployed, and each request is routed to a small subset of the most suitable experts based on a gating mechanism. This innovation enables MoE models to scale model parameters without incurring extra inference overhead, since only a fraction of parameters are activated for each request. However, this mechanism also introduces dynamic randomness, since expert selection is unknown until gating is completed, posing new challenges for serving systems.

\subsection{Prior MoE Serving Systems}
The MoE mechanism constitutes the primary source of data movement overhead in modern serving systems. As illustrated in ~\autoref{fig:1-breakdown}, take DeepSeek V3 as an example,
MoE-related data movement (MoE All-to-All and MoE Weights) dominates the overhead across different serving configurations, accounting for 60\%-90\% of total latency under 4K sequence length.
To address this , existing research has developed numerous system-level solutions targeting different performance and cost objectives. 
Edge systems like MoE-Lightning~\cite{moelightning} and CoServe~\cite{suo2025coserve} employ CPU memory offloading techniques to address GPU memory capacity constraints, while cloud systems such as Comet~\cite{zhang2025comet} and MegaScale-Infer~\cite{zhu2025megascale-infer} target multi-GPU systems and address GPU-GPU communications in MoE for higher throughput. 
Novel hardware architectures like Duplex~\cite{yun2024duplex} explore processing-in-memory to accelerate data movement in MoE LLMs. 

However, these prior studies employ a \emph{system-centric} methodology when optimizing for MoE LLMs. Namely, they inherently focus on a specific platform and the corresponding data movement patterns of MoE in such platform (e.g., CPU-GPU, multi-GPU, ML accelerators). As a result, they propose deployment-specific optimizations that may not generalize across different serving platforms, and their insights are often a slice of the overall inherent patterns in MoE LLMs.

In this work, we flip the process and adopt a \emph{model-centric} strategy by conducting system-independent profiling to extract \emph{system-agnostic} insights about MoE data movement patterns. These insights are therefore broadly applicable across various platforms, providing a foundation for optimization strategies that transcend specific system implementations.



\begin{figure}[tbp]
    \centering
    \includegraphics[width=0.49\textwidth]{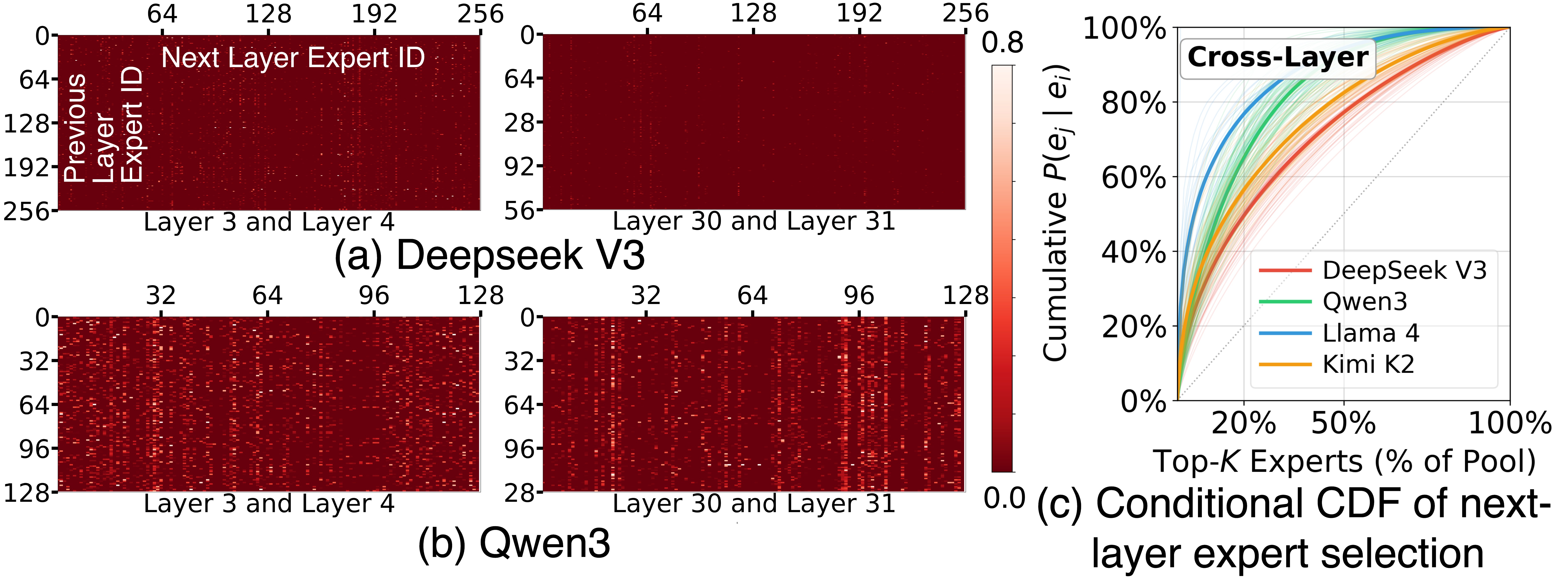}
    \caption{Cross-layer expert correlation. (a, b) Joint co-activation heatmaps between layers $N$ and $N{+}1$ in DeepSeek-V3 and Qwen3. (c) Conditional CDF 
    $P(e_j\!\mid\!e_i)$ 
    for each layer's top-1 expert.}
    \label{fig:profile-layer-level}
\end{figure}

\begin{figure*}[ht!]
    \centering
    \includegraphics[width=0.99\textwidth]{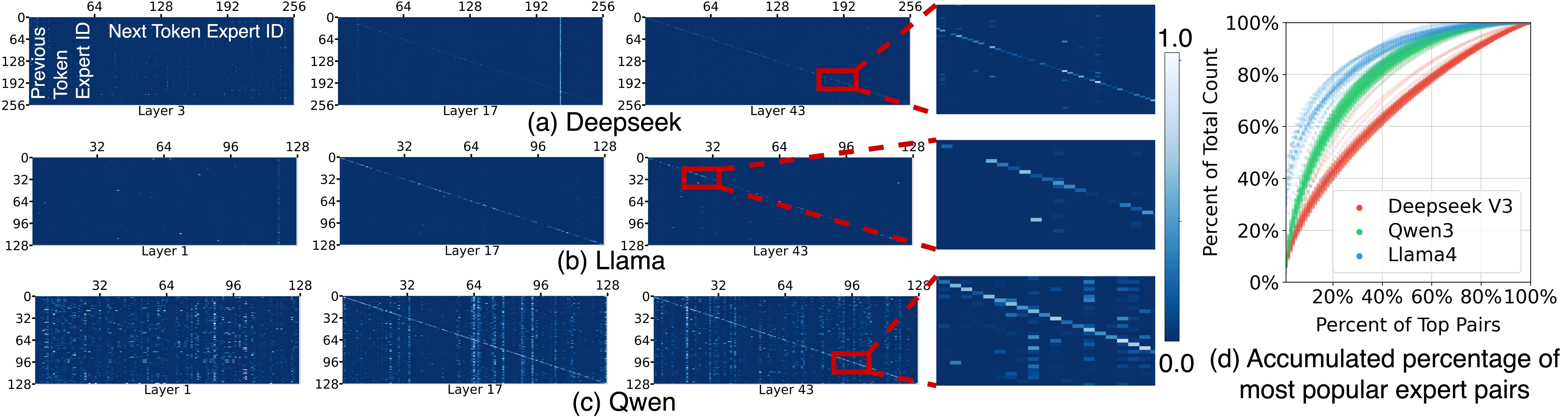}
\caption{Cross-token expert correlation. (a, b, c) Joint co-activation heatmaps between tokens $t$ and $t{+}1$ in DeepSeek-V3, Llama 4, and Qwen3. (d) Conditional CDF for each layer's top-1 expert: top 20\% of the next-token expertalready covers most of 
probability mass.}
    \label{fig:profile-token-level}
\end{figure*}

\begin{figure}[ht!]
    \centering
    \includegraphics[width=0.49\textwidth]{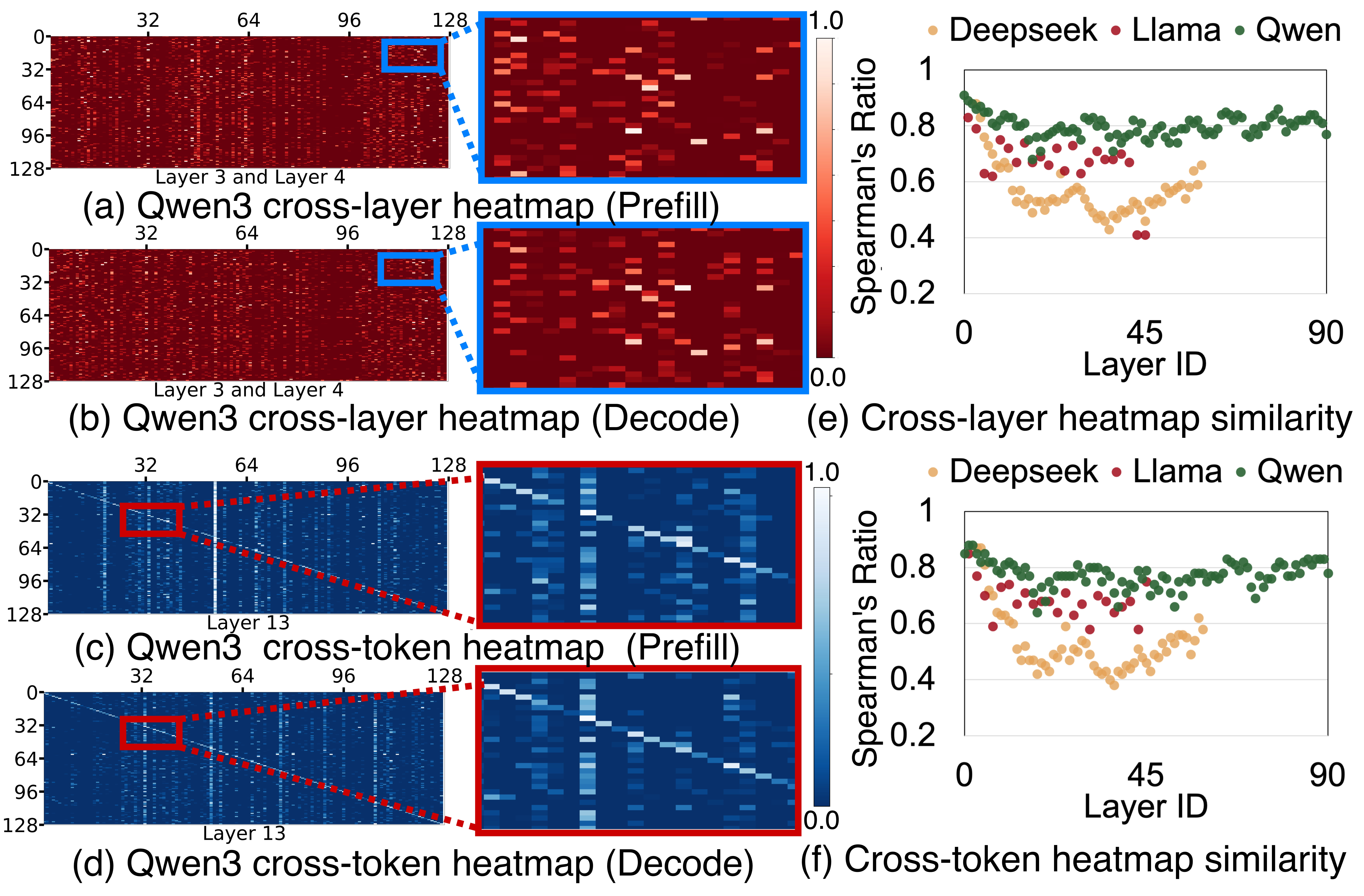}
    \caption{Expert activation patterns remain consistent across prefill and decode stages for both (a, b) cross-layer heatmap and (c, d) cross-token heatmaps. Spearman's ratio quantified in (e, f) shows a strong relation ($\geq0.7$).}
    \label{fig:profile-stage-level}
\end{figure}

\section{MoE Profiling and System Insights}
In this section, we conduct a data-movement-centric profiling of the expert selection behavior in four state-of-the-art MoE models: Deepseek V3 (671B), Llama4-Maverick-128E (402B), Qwen3-235B (235B), and Kimi K2 (1000B). 
All results are averaged over more than 24,000 requests.

\subsection{Categorization Methodology}

As in \autoref{fig:2_intro}, we categorize MoE expert selection profiling results into two categories:
\emph{temporal} and \emph{spatial} relations. 

\textbf{Temporal relations} capture time-dependent expert selection patterns where current choices inform future selections. These patterns enable \emph{single-unit strategies} that optimize data movement for individual units through prefetching, caching, and data migration. For example, in multi-chiplet GPU systems, caching experts in local DRAM after remote fetches significantly reduces inter-unit communication. To exploit temporal predictability, we analyze expert selection at multiple time scales shown in \autoref{fig:2_intro}(a): \emph{layer-level}, \emph{token-level}, and \emph{prefill-decode-level} patterns.

\textbf{Spatial relations} capture how expert activations are distributed across compute units within a given time window. This distributional information enables \emph{multi-unit strategies} that optimize expert placement and workload balancing across the system, reducing data movement and preventing bottlenecks. We classify spatial relations into \emph{single-expert activation imbalance} and \emph{expert-pair co-activation affinity} as shown in \autoref{fig:2_intro}(b), and investigate how task types influence these patterns to inform system-level optimization.

\subsection{Temporal Relations}

As shown in ~\autoref{fig:2_intro}(a), we classify the temporal relations of expert selection into three categories, arranged in order of increasing time scale.
At the layer level, we examine the relationship between two adjacent model layers.
At the token level, we focus on the same model layer across two adjacent tokens.
At the stage level, we analyze the relationship between the prefill stage and the decode stage.

\subsubsection{\textbf{Layer-Level Correlation}}
\label{subsubsec: layer-level}
\textbf{\emph{(Ob1)}}\\
As shown in \autoref{fig:profile-layer-level} (a) and (b), we present heatmaps for Deepseek and Qwen illustrating expert selection relationships across adjacent layers. Each pixel in the heatmap displays the conditional probability of selecting expert $j$ in the next layer given that expert $i$ was activated in the previous layer, with bright colors indicating higher probabilities.

The heatmaps reveal clear cross-layer correlations with white dots highlighting specific expert pairs with significantly higher selection probabilities across adjacent layers. However, correlation patterns vary across layers within the same model and differ between models due to architectural variations. For instance, patterns between layers 3-4 differ from those between layers 30-31. Qwen3's notably brighter heatmap indicates stronger cross-layer correlations than Deepseek.
Beyond the white dots, there are also consistent bright vertical lines, suggesting certain experts are frequently chosen regardless of previous layer selections. These patterns indicate generally popular experts, analyzed further in Sec. \ref{subsubsec: single-expert}. 

To quantify these relationships, we analyze the conditional CDF $P(e_j \mid e_i)$ in \autoref{fig:profile-layer-level}(c): the top 20\% of
next-layer candidates already cover 50\%, 65\%, 77\%, and 56\% of the conditional probability mass for DeepSeek-V3, Qwen3, Llama~4\footnote{Llama~4 inserts dense FFN layers between MoE
layers, so we pair adjacent MoE layers ($N$ and $N{+}2$).}, and Kimi K2, respectively.
This reveals strong, model-dependent cross-layer correlations, with Llama4 showing the strongest effect and Deepseek the weakest.
    
\subsubsection{\textbf{Token-Level Correlation}}
\label{subsubsec: token-level}
\textbf{\emph{(Ob2)}}\\
We examine expert selection relations for the same layer between adjacent tokens in \autoref{fig:profile-token-level}. 
Each pixel in the heatmap displays the conditional probability of selecting expert $j$ in the next token given that expert $i$ was activated in the previous token, with bright colors indicating higher probabilities.

Similar to layer-level patterns, cross-token heatmaps exhibit white dots, bright vertical lines, and variation across layers and models, indicating correlations between adjacent tokens. However, token-level relations reveal a common pattern appearing across all models: the bright diagonal line that indicates the tendency to select the same expert across adjacent tokens. 
This diagonal pattern emerges predominantly in higher layers (17 and 43) but not in lower layers (1 and 3), regardless of models.

We apply the same conditional-CDF analysis to the token-level relation. As shown in \autoref{fig:profile-token-level}(d), the top 20\% of next-token expert candidates  cover 47\%, 62\%, 80\%, and 53\% of the cumulative conditional probability in DeepSeek-V3,
Qwen3, Llama~4, and Kimi K2, respectively, averaged across all MoE layers. 
The correlation is again strongest in Llama~4 and
weakest in DeepSeek-V3.

\subsubsection{\textbf{Prefill-Decode-Level Correlation}}
\label{subsubsec: stage-level}
\textbf{\emph{(Ob3)}}\\
Building on the layer-level and token-level relations, we observe notable similarities in expert selection patterns between prefill and decode stages.
Comparing heatmaps across stages in \autoref{fig:profile-stage-level}(a)(b)(c)(d), we find similar distributions of bright dots, indicating that expert pair heatmap during prefill and decode shares similarities.
This cross-stage consistency suggests us that the prefill-collected information can guide initial decode steps until sufficient decode data accumulates.

To quantify this similarity, we compute Spearman's ratio ($\rho$) across all model layers, comparing prefill and decode heatmaps. Spearman's Ratio $\rho$ measures monotonic relationships between variables, ranging from $-1$ (perfect negative correlation) to $1$ (perfect positive correlation). Generally, $|\rho| > 0.7$ indicates strong correlation, $0.4 < |\rho| \leq 0.7$ indicates moderate correlation, and $|\rho| \leq 0.4$ suggests weak correlation~\cite{spearman}.
The results in \autoref{fig:profile-stage-level}(e)(f) show that most layers demonstrate strong correlation, while a few show moderate correlation. This makes it possible to predict decode-stage expert selection with prefill-stage data.

\begin{figure}[t]
    \centering
    \includegraphics[width=0.49\textwidth]{fig/temporal_stage_single_exp.pdf}
    \caption{(a) Expert frequency distributions between prefill and decode
    stages exhibit similarity. (b) The most popular experts in the two stages
    overlap substantially. (c) All models show a high Spearman correlation
    between prefill and decode expert frequencies.}
    \label{fig:stage-level_single_exp}
\end{figure}

Beyond expert-pair heatmaps, we also identify prefill-to-decode correlation at the single-expert frequency level. As shown in~\autoref{fig:stage-level_single_exp}(a), the frequency distributions of prefill and decode stages are substantially similar, though some discrepancies exist among low-frequency experts. To examine the most popular experts, we report the overlap rate of top experts between stages in~\autoref{fig:stage-level_single_exp}(b): the top-5 prefill experts cover around 60\% of the top-5 decode experts, rising to 75\% and 90\% for top-10 and top-20, respectively. This indicates that prefill information can help predict the hottest decode experts. The cross-model Spearman correlation in~\autoref{fig:stage-level_single_exp}(c) confirms this relationship holds across all four models.

\subsubsection{\textbf{System Insights from Temporal Relation}}
\label{subsubsec: temporal insights}

The observed temporal relations in expert selection motivate us to design fine-grained, dynamic strategies on every single unit to reduce data movement.
For example, when expert weights are read from remote memory, such as remote DRAM in multi-chiplet systems, or CXL extension memory in memory-disaggregated systems, caching, migration, and prefetching strategies can be deployed to reduce data movement.

\begin{tcolorbox}[label=box:insight1, colframe=black, colback=white, coltitle=black, boxrule=0.5mm, left=0.5em, right=0.5em, top=0.5em, bottom=0.5em]
\ding{72}\emph{\textbf{Insight 1: Prefill-data-driven prediction (\hyperref[subsubsec: layer-level]{\textcolor{magenta}{\underline{Ob3}}}).} Leverage the expert selection trace from the prefill stage to predict expert selection during the decoding stage.}
\end{tcolorbox}

Empirical analysis shows that expert selection patterns during prefill exhibit strong similarity to those during decode. Thus, expert selection information collected in the prefill phase can serve as a valuable reference for predicting decode-phase selections, particularly at the beginning of decoding when only a few tokens have been generated and historical context is scarce. Our~\autoref{sec: case study 2} demonstrates how prefill information can guide expert placement during decode. This is especially relevant in modern PD-disaggregated serving systems, where the prefill and decode stages execute on separate machines. 

\begin{tcolorbox}[label=box:insight2, colframe=black, colback=white, coltitle=black, boxrule=0.5mm, left=0.5em, right=0.5em, top=0.5em, bottom=0.5em]
\ding{72}\emph{\textbf{Insight 2: Cross-hierarchy memory management (\hyperref[subsubsec: layer-level]{\textcolor{magenta}{\underline{Ob1}}}, \hyperref[subsubsec: stage-level]{\textcolor{magenta}{\underline{Ob2}}}).} Token- and layer-level temporal relations enable dynamic expert prefetching and caching across memory hierarchies.}
\end{tcolorbox}

Layer-level and token-level temporal relations are similar in definition but differ in reuse distance, making them suitable for different levels of the memory hierarchy. Layer-level relations exhibit short reuse distances because consecutive MoE layers execute in immediate succession, while token-level relations incur longer reuse distances because a new token is generated only after traversing all layers.

This maps naturally onto the multi-level memory hierarchies in modern serving systems. For example, in multi-chiplet architectures, each die contains both an LLC and local DRAM, forming a two-tier hierarchy. The faster but smaller LLC is well-suited to managing experts with short reuse distances (layer-level), while the larger local DRAM accommodates experts with longer reuse distances (token-level). 
Accordingly, we can leverage layer-level relations for LLC management and token-level relations for DRAM management.

This principle generalizes to other system configurations: CXL-based systems with local DRAM and remote CXL memory, SSD offloading systems with DRAM and flash storage, and PIM systems with local and remote DRAM dies. In each case, layer-level relations guide the faster memory tier and token-level relations guide the slower one.

\begin{figure}[t!]
    \centering
    \includegraphics[width=0.48\textwidth]{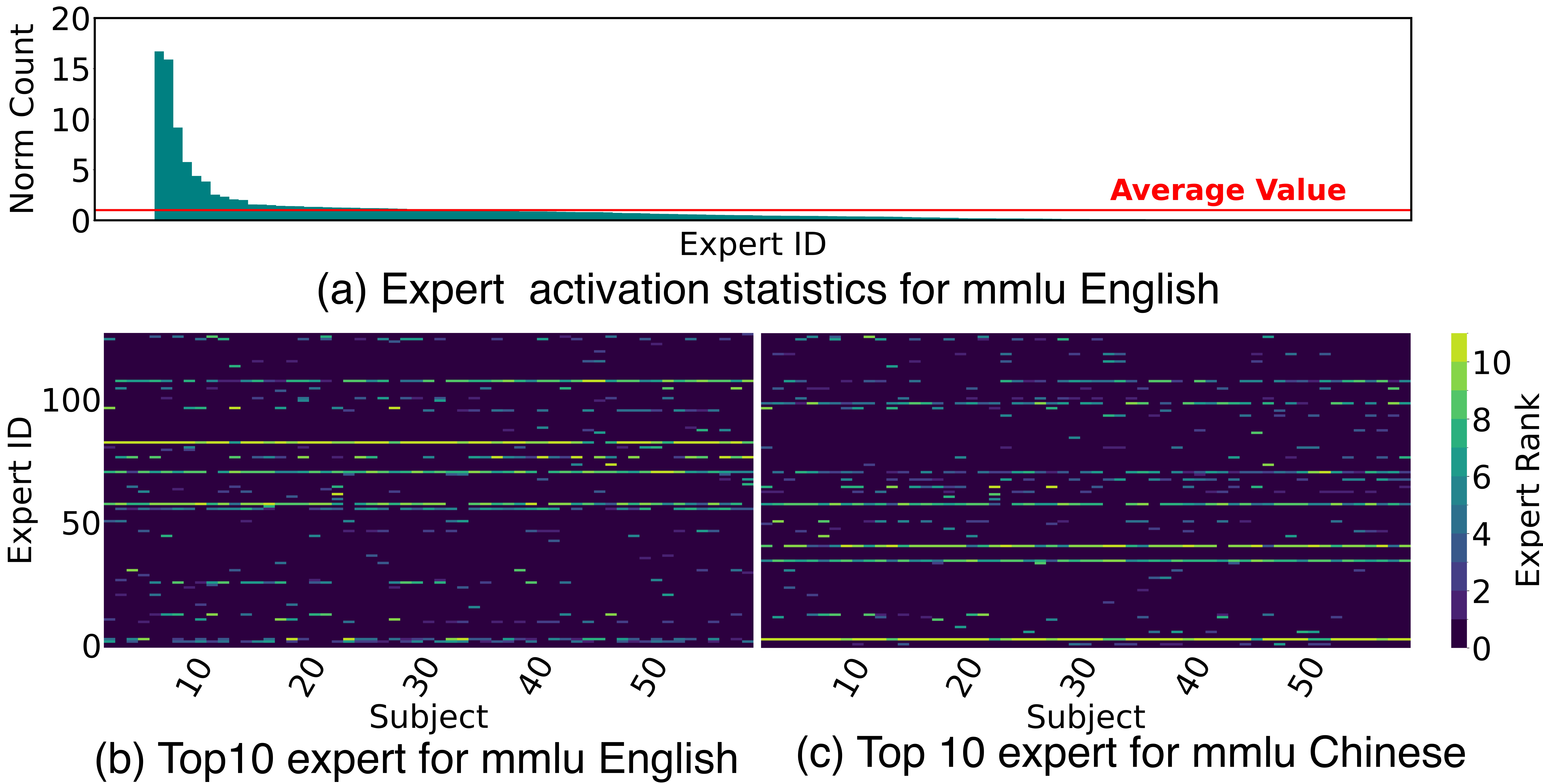}
    \caption{Single-expert spatial relation analysis of Llama4 layer 7 shows: (a) non-uniform expert activation distribution; (b) expert selection strongly correlates with task type; (c) expert activation patterns shift significantly when language changes while content remains identical.}
    \label{fig:profile-single-expert}
\end{figure}

\subsection{Spatial Relation}
\label{subsec: spatial relation}
As shown in \autoref{fig:2_intro}(b), we analyze spatial patterns in expert selection for both single-expert activation imbalance and expert pair co-activation affinity. 
For single experts, we examine statistical skewness and the factors affecting each expert's activation. For expert pairs, we analyze co-activation properties across all two-expert combinations.

\subsubsection{\textbf{Single Expert Activation Imbalance}}
\label{subsubsec: single-expert}
\textbf{\emph{(Ob4)}}\\
We examine expert selection frequency at each layer, presenting results for layer 7 of Llama4 in~\autoref{fig:profile-single-expert}. 
We observe pronounced skewness where a subset of experts is activated over 16 times more frequently than average. This workload imbalance suggests system designs should duplicate or decentralize frequently used experts.

To investigate selection patterns across different tasks, we analyze all 57 MMLU subjects spanning diverse fields, including biology, history, and math, etc~\cite{mmlu}. 
\autoref{fig:profile-single-expert}(b) shows the top 10 most popular experts for each subject. Horizontal bright lines indicate certain experts are consistently activated regardless of subject, while remaining popular experts vary significantly between subjects, demonstrating both overlap and distinction in task-based expert selection.

We further examine task impact using the Chinese version of MMLU in MMLU Pro~\cite{mmlu_pro} with identical questions but different languages. \autoref{fig:profile-single-expert}(c) reveals distinctly different patterns: although 5-6 experts remain popular across subjects, only two overlap with English MMLU's most frequently selected experts. This confirms that task characteristics, including language, significantly influence expert selection, enabling task-aware serving systems that optimize expert distribution to balance workloads and reduce data movement.

\subsubsection{\textbf{Expert Pair Co-activation Affinify}}
\label{subsubsec: expert pair}
\begin{figure}[t!]
    \centering
    \includegraphics[width=0.49\textwidth]{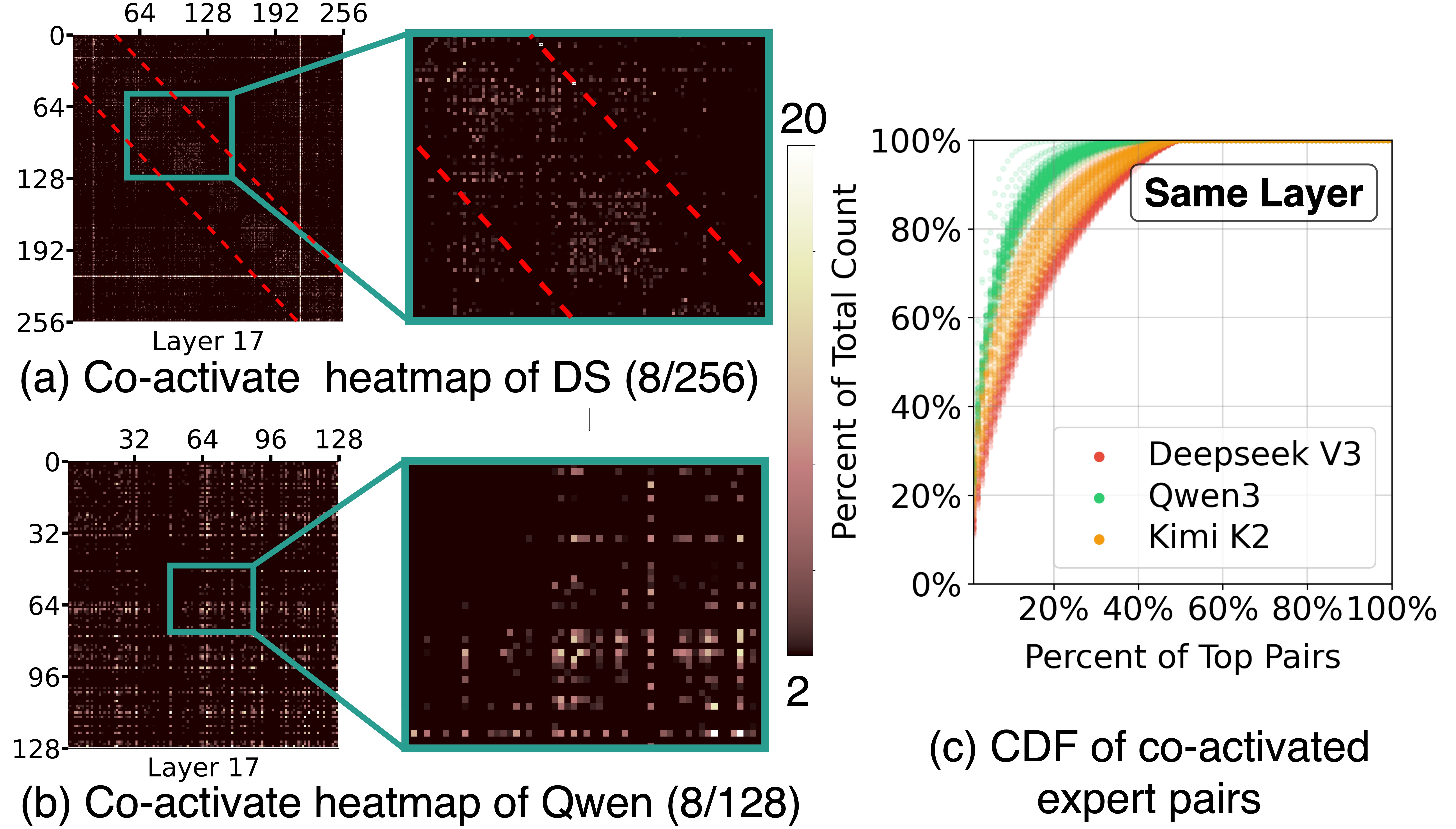}
    \caption{
Expert-pair co-activation affinity. (a)(b) Heatmaps for DeepSeek and Qwen. (c) CDF of co-activated expert pairs across all layers: a small fraction of expert pairs accounts for the majority of co-activations.}
    \label{fig:profile-expert pair}
\end{figure}

\textbf{\emph{(Ob5)}}\\
Beyond single expert patterns, we observe spatial relations for expert pairs where certain experts are more likely to be co-activated. We present co-activation heatmaps in \autoref{fig:profile-expert pair}(a)(b), where both axes indicate expert IDs. Each pixel represents an expert pair with values showing co-activation frequency normalized by theoretical random selection probability: $p = \frac{2}{n(n-1)}$, where $n$ is the number of experts.

Bright dots appear with probabilities 20-40 times higher than theoretical values, indicating strong co-activation tendencies. All heatmaps exhibit central symmetry since expert pair $(i, j)$ equals $(j, i)$. In Deepseek's heatmap \autoref{fig:profile-expert pair}(a), frequently activated pairs lie between red lines forming bright squares, reflecting Deepseek's routing restriction where tokens are routed only to adjacent nodes to reduce communication overhead. This suggests the potential of separating co-activated expert pairs to balance workload.

We quantify this relation by in \autoref{fig:profile-expert pair}(c). 
The top 10\% of expert pairs account for 60-80\% of total activations, indicating strong skewness. 
This suggests the potential for separating co-activated expert pairs to balance the workload.
We only analyze Deepseek and Qwen since Llama selects one expert per MoE layer, eliminating co-activation relations.

\subsubsection{\textbf{System Insights from Spatial Relation}}
Spatial Relation enables coarse-grained, static strategies to address workload imbalance across the system. These strategies could be applied at system startup or during periodic redistribution (e.g., every 10 minutes) through appropriate task distribution.

\begin{tcolorbox}[label=box:insight3, colframe=black, colback=white, coltitle=black, boxrule=0.5mm, left=0.2em, right=0.2em, top=0.5em, bottom=0.5em]
\ding{72}\emph{\textbf{Insight 3: Expert-placement-aware workload distribution (\hyperref[subsubsec: single-expert]{\textcolor{magenta}{\underline{Ob4}}}, \hyperref[subsubsec: expert pair]{\textcolor{magenta}{\underline{Ob5}}}).} Employ expert placement information to design workload-balanced task distribution strategies.}
\end{tcolorbox}


Expert placement in serving systems can change dynamically due to expert migration strategies. Therefore, when allocating workload to system units, expert placement should be considered for better workload balance. Besides, the design space for task allocation could be enlarged with emerging new systems. 
Traditional multi-GPU systems tend to allocate experts to local GPUs to avoid cross-unit communication. 
However, in multi-chiplet GPUs, we can consider allocating tasks to remote dies for better workload balance as inter-unit communication becomes faster.

\begin{tcolorbox}[label=box:insight4, colframe=black, colback=white, coltitle=black, boxrule=0.5mm, left=0.5em, right=0.5em, top=0.5em, bottom=0.5em]
\ding{72}\emph{\textbf{Insight 4: Popular expert decentralization (\hyperref[subsubsec: single-expert]{\textcolor{magenta}{\underline{Ob4}}}).} Duplicate or decentralize frequently used experts to balance workloads.} 
\end{tcolorbox}

Expert skewness causes workload imbalance and suboptimal resource utilization. Duplicating popular experts across multiple compute units distributes load more evenly. Additionally, avoiding co-location of highly popular experts in the same unit further enhances workload balance.

\begin{tcolorbox}[label=box:insight5, colframe=black, colback=white, coltitle=black, boxrule=0.5mm, left=0.5em, right=0.5em, top=0.5em, bottom=0.5em]
\ding{72}\emph{\textbf{Insight 5: Expert-pair separation (\hyperref[subsubsec: expert pair]{\textcolor{magenta}{\underline{Ob5}}}).} 
Separate frequently co-activated expert pairs to maximize parallelism.}
\end{tcolorbox}

Certain experts are frequently activated simultaneously, exhibiting strong co-activation patterns. Assigning these co-activated expert pairs to different compute units maximizes hardware parallelism and prevents workload concentration on specific units. However, separation also introduces cross-unit communication overhead. The effectiveness depends on system topology and batch size, requiring careful trade-off between parallelism benefits and communication costs.

\begin{tcolorbox}[label=box:insight6, colframe=black, colback=white, coltitle=black, boxrule=0.5mm, left=0.5em, right=0.5em, top=0.5em, bottom=0.5em]
\ding{72}\emph{\textbf{Insight 6: Workload-aware serving system (\hyperref[subsubsec: single-expert]{\textcolor{magenta}{\underline{Ob4}}}).}
Leverage the workload information, like task type and language, to make expert migration prior to serving.} 
\end{tcolorbox}

Hot experts vary by task and language. English queries, for instance, activate different expert subsets than Chinese queries. Providing task metadata during serving enables proactive expert placement: when workloads are predominantly English, systems can pre-duplicate or reassign English-relevant experts, reducing communication and balancing loads. This task-to-expert mapping requires only one-time offline profiling per model and can be reused throughout deployment, making the approach practical and efficient.

\section{Case Study 1: Wafer-scale GPU Architecture Design for MoE Serving}

In this section, we adopt future GPU architecture design as a use case to validate our proposed insights. We follow \hyperref[box:insight3]{\textcolor{magenta}{\underline{Insight 3}}} to design a task allocation algorithm and leverage the temporal relation insights (\hyperref[box:insight1]{\textcolor{magenta}{\underline{Insight 1}}} and \hyperref[box:insight2]{\textcolor{magenta}{\underline{Insight 2}}}) to build a data-driven predictor. We also make slight architectural modifications to support the proposed strategies.

\subsection{Trend of Future GPU Architecture}

GPU vendors are increasingly adopting multi-chiplet architectures to overcome single-die performance limitations. As Moore's Law approaches its limits~\cite{lundstrom2003moore} and single-die size remains constrained by photomask dimensions (800-1,000 mm$^2$), advanced packaging technologies like TSMC's CoWoS~\cite{hu2023cowos}, Samsung's X-Cube~\cite{xcube}, and Intel's EMIB~\cite{emib} enable multiple chiplets within a single package. Leading vendors have adopted such designs: AMD's MI300~\cite{mi300} integrates eight compute chiplets, NVIDIA's Blackwell features two chiplets~\cite{blackwell}, and upcoming Rubin expects four~\cite{rubin}.

\begin{figure*}[t]
    \centering
    \includegraphics[width=0.99\textwidth]{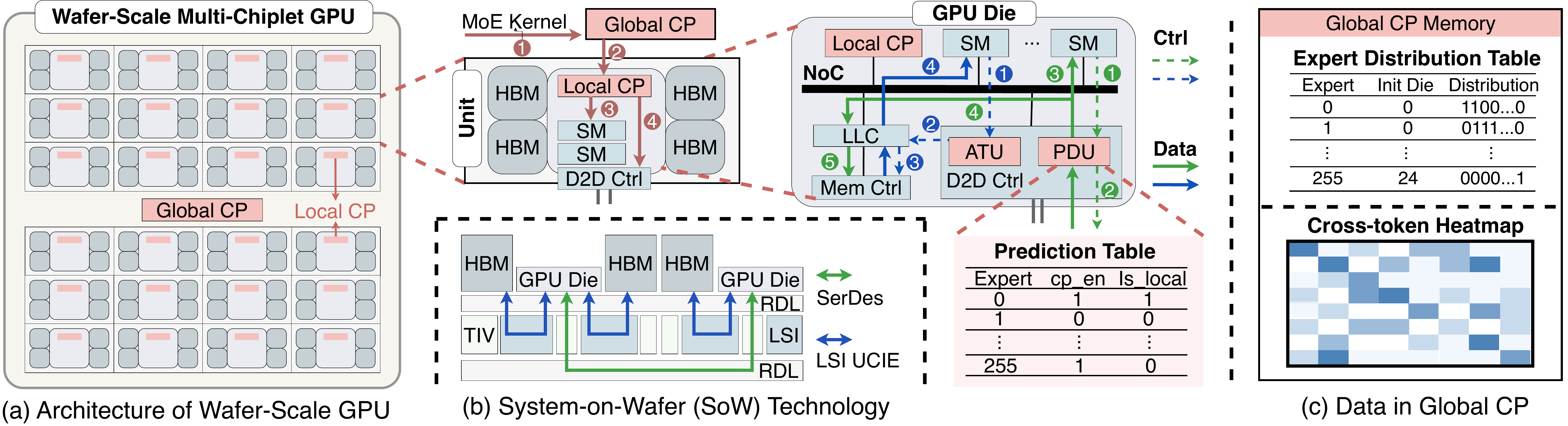}
    \caption{(a) Wafer-scale multi-chiplet GPU architecture with additional units highlighted in orange. (b) SoW (System-on-Wafer) technology structure. (c) Data format in the Global Command Processor for our proposed task distribution strategy.}
    \label{fig:5_architecture}
\end{figure*}

This trend is evolving toward wafer-scale systems~\cite{cowos2}. TSMC's System-on-Wafer (SoW) technology accommodates up to 24 compute dies and 96 HBM dies on a single wafer, exceeding 200,000 mm$^2$~\cite{sow-x}. 
As shown in \autoref{fig:5_architecture}(a), a typical wafer-scale multi-chiplet GPU consists of multiple units, each containing a GPU die and several HBM dies interconnected in a mesh topology. Such systems contain over 3 TB of HBM and PFLOPS-level computing power, supporting extremely large models and batch sizes.

The TSMC SoW technology shown in \autoref{fig:5_architecture}(b) connects each GPU die to local HBM dies through local-silicon interconnects (LSI)~\cite{lsi}, with adjacent GPU dies communicating via LSI vertically or XSR SerDes links horizontally.
Although LSI and SerDes both provide terabit-level bandwidth, inter-GPU-die communication remains the primary bottleneck. 
Remote data access requires multiple hops across die-to-die links, resulting in high latency. 
Simultaneous remote HBM access by multiple dies creates bandwidth contention and traffic congestion, further degrading performance.

\subsection{Background on Wafer-scale GPU Programming model}
The programming model of future wafer-scale chip remains an open question, but two major candidates have emerged: the multi-GPU-like and the single-GPU-like programming model.

\textbf{Multi-GPU-like programming model.}
WSC-LLM~\cite{xu2025wsc} and MoEntwine~\cite{tang2025moentwine} adopt
a multi-GPU-like programming model that exposes the entire wafer as a
multi-GPU system.
Programmers can program the wafer similarly to a conventional
multi-GPU system, with the key difference being the 2D mesh topology where each die communicates directly only with its neighbors.
While this approach offers fine-grained control over individual dies and enables flexible software strategies, it diverges from current industry trends.
For instance, although Blackwell and Rubin both integrate two compute
dies, NVIDIA exposes no toolchain for fine-grained die-level control.
Multi-Instance GPU (MIG) can partition a multi-die GPU into independent
GPU instances, making each die act independently, but the high-speed D2D link is disabled in this
mode~\cite{joshi2026mig}, forcing inter-die communication through the
10-100$\times$ slower NVLink or PCIe and negating the benefit of multi-die
packaging.
Therefore, extending this programming model to wafer-scale GPUs would require substantial architectural changes to current GPU designs, including redesigning the D2D/C2C controller workflow and distributed LLC structure, making it infeasible in the near term.

\textbf{Single-GPU-like programming model.} 
HDPAT~\cite{xuhdpat}, Hecton~\cite{huang2024hecaton}, and our work adopt a single-GPU-like programming model that exposes the entire wafer as a unified GPU, fully abstracting the multi-die topology and data placement from software so that the programming experience is identical to that of a monolithic GPU. 
We adopt this model for two key reasons. First, it aligns with commercial multi-chiplet GPUs such as Blackwell and Rubin, ensuring practical industry relevance. 
Second, it eliminates the programming complexity of explicit inter-die communication management through libraries such as NCCL or NVSHMEM. 
However, this programming simplicity shifts the optimization burden entirely to the hardware. Given the inherently distributed architecture, local versus remote data access costs vary by up to 15$\times$, yet the abstraction prevents programmers from controlling cross-die data movement. Consequently, architecture-level optimizations become critical to achieve high hardware utilization.

\subsection{Challenges of serving MoE with future GPUs}
Unlike current multi-GPU systems, wafer-scale GPUs can fit entire MoE models on a single chip and support batch sizes over 10,000. However, current GPU architectures introduce two key limitations for such large-scale chips.

\textbf{Simplistic Task Allocation.} Current GPUs integrate a CPU in their SoC to serve as a command processor and allocate tasks to all SMs. However, the traditional command processors treat all SMs equally, ignoring their physical locations and data placement~\cite{cp1, cp2}. This oblivious task-to-SM assignment generates excessive D2D traffic and ignores MoE expert selection skewness, leading to poor utilization when most dies remain idle while others become overloaded.

\textbf{Inadequate Local HBM Management.} Current GPUs treat all HBM dies as uniform memory space, but wafer-scale GPUs connect each compute die directly to local HBM, where access is significantly faster than a remote HBM. 
Frequently accessed experts in remote HBM could be cached locally to minimize D2D traffic, but current GPUs do not distinguish between local and remote HBM and therefore generate unnecessary traffic.

\subsection{Motivation and Insights}
To address these challenges, we propose two strategies with architectural support. 
First, based on
\hyperref[box:insight3]{\textcolor{magenta}{\underline{Insight 3}}}
that identifies the need for expert-placement-aware task distribution, we propose an intelligent task allocation algorithm with a multi-level, data-placement-aware command processor architecture. 
This approach considers expert placement and selection skewness across dies, enabling dynamic task allocation that minimizes D2D traffic while balancing workload.

Second, leveraging \hyperref[box:insight1]{\textcolor{magenta}{\underline{Insight 1}}} and \hyperref[box:insight2]{\textcolor{magenta}{\underline{Insight 2}}} that reveal the predictability behind expert selection across different timescales, we introduce a data-driven predictor with hardware-managed HBM architecture. 
Local HBM caches frequently accessed experts from remote dies, while a lightweight predictor analyzes selection patterns to estimate future needs, caching predicted experts locally to reduce D2D traffic.

To implement these two strategies under a single-GPU-like programming model, we a few architectural extensions to the GPU architecture. 
If future programming models evolve toward multi-GPU-like abstractions with finer-grained control over each die, these strategies could alternatively be realized at the system level without any architectural modification.

\subsection{Architecture Design}
\subsubsection{Architecture Overview}
To support our proposed predictor and task allocation algorithm, we implement two architectural modifications with minimal overhead, as illustrated in \autoref{fig:5_architecture}(a). These changes, highlighted in orange, consist of an enhanced Command Processor structure and an extended D2D controller design.

First, we redesign the Command Processor (CP) with a two-level hierarchical structure: a Global CP at the wafer level and Local CPs within each die. The Global CP maintains system-wide expert selection and placement information for intelligent resource management.
Second, we extend the D2D controller with an Address Translation Unit (ATU) and a Prediction Unit (PDU). The ATU translates remote HBM addresses to local addresses when remote data is duplicated, while the PDU identifies important remote data requiring duplication. These enable autonomous caching of remote data in local HBM and intelligent redirection of data requests, reducing inter-die communication overhead.

\subsubsection{Key Data Structures}
There are two key data structures: Global CP data and the PDU prediction table.

\noindent\underline{Global CP Data Structures: } As shown in \autoref{fig:5_architecture}(c), the Global CP maintains two structures. The expert distribution table stores each expert's initial die ID and distribution status as an $n$-bit binary code, where each bit indicates expert presence on the corresponding die. The cross-token heatmap records expert activation patterns over time, providing historical data for prediction generation.

\noindent\underline{PDU Prediction Table: } Each PDU stores a prediction table with two key fields per expert: the cp\_en bit indicating whether the expert should be cached locally (calculated by Global CP and transferred to each die), and the is\_local bit tracking whether the expert is already cached in local HBM.

\subsubsection{Workflow During Kernel Launch}
When a new kernel launches ({\small \circlednumber{myred}{1}}), the Global CP runs our task allocation algorithm to split the MoE kernel into per-die sub-kernels and executes the predictor to generate duplication guidance (cp\_en field is PDU). The Global CP then sends sub-kernels and prediction information to local CPs ({\small \circlednumber{myred}{2}}). Each local CP allocates tasks to its SMs ({\small \circlednumber{myred}{3}}) and configures the prediction table in the D2D controller for local HBM management ({\small \circlednumber{myred}{4}}). After computation, local CPs collect expert duplication statistics and send them to Global CP to update expert distribution information.

\subsubsection{Dataflow for Remote Data Access}
We integrate ATU and PDU into the D2D controller to support hardware-managed HBM by modifying the remote data access flow. 
With these two units, our architecture automatically duplicates important remote data in local HBM, with green lines representing non-duplicated data reads and blue lines representing duplicated data reads, as shown in \autoref{fig:5_architecture}(a).

\noindent\underline{Remote Data Read (Non-duplicated): } When an SM reads remote data not in local HBM ({\small \circlednumber{mygreen}{1}}), the D2D controller routes the request conventionally ({\small \circlednumber{mygreen}{2}}). 
Upon return, the PDU checks the Prediction Table to make a duplication decision and sends data to the SM regardless of the decision ({\small \circlednumber{mygreen}{3}}). 
If duplication is required, the PDU writes to LLC and local HBM ({\small \circlednumber{mygreen}{4}}, {\small \circlednumber{mygreen}{5}}), updates the address into ATU, and sets the is\_local bit in PDU's Prediction Table to 1.

\noindent\underline{Local Data Read (Duplicated): } When an SM reads remote data already cached in local HBM ({\small \circlednumber{myblue}{1}}), the ATU translates the remote address to a local address and redirects the request to LLC ({\small \circlednumber{myblue}{2}}). 
The LLC and memory controller then retrieve data and send it back to the SM({\small \circlednumber{myblue}{3}}, {\small \circlednumber{myblue}{4}}).

\subsubsection{Algorithm Design}
This subsection presents our task allocation algorithm and data-driven predictor, both implemented by the global CP.

\noindent\underline{Task Allocation Algorithm: }
Since accurate task distribution is NP-hard, we propose two heuristic mechanisms: a candidate mechanism that reduces the number of dies to consider and a block-granularity distribution mechanism that searches for approximate solutions among candidates.

This algorithm splits MoE kernel computation into sub-tasks for individual dies based on expert selection and distribution information. 
As shown in ~\autoref{alg:task_allo}, the input $expert\_reqs\_dict$ contains the number of requests belonging to each expert, while $expert\_die\_map$ provides dynamic expert distribution information from the Expert Distribution Table in \autoref{fig:5_architecture}(c), indicating where each expert is stored.

\SetKwFunction{GenCandidateList}{GenCandidateList}
\SetKwFunction{Sort}{Sort}
\SetKwFunction{CostModel}{CostModel}
\SetKwFunction{Argmin}{Argmin}
\SetKwFunction{Update}{Update}
\SetKwFunction{MergeTasks}{MergeTasks}
\SetKwFunction{FindNearDies}{FindNearDies}





\begin{algorithm}[t]
\footnotesize
\LinesNumbered
\caption{Task Allocation Algorithm}
\label{alg:task_allo}
\KwIn{$expert\_reqs\_dict$, $expert\_die\_map$}
\KwOut{$allo\_plan$}

\textbf{Initialize} the workload of each die: $load\_per\_die$\;
\Sort{$expert\_reqs\_dict$, by $req\_num$ ascending}\;
\For{$(expert\_id, req\_num)$ in $expert\_reqs\_dict$}{
    $candi\_list \gets$ \GenCandidateList{$expert\_id$, $dis\!=\!1$}\;
    $candi\_list \gets$ \Sort{$candi\_list$, $i \mapsto load\_per\_die[i]$}
    \While{$req\_num > 0$}{
        $costs\_of\_dies \gets$ \CostModel{$candi\_list$}\;
        $target\_die \gets$ \Argmin{$costs\_of\_dies$}\;
        $allo\_plan$.append([$expert\_id$, $target\_die$, $req\_blk$])\;
        \Update{$load\_per\_die$}\;
        $req\_num \gets req\_num - req\_blk$\;
    }
    $allo\_plan \gets$ \MergeTasks{$allo\_plan$}\;
}
\Return $allo\_plan$\;

\SetKwProg{Fn}{Function}{:}{}
\Fn{\GenCandidateList{$expert\_id$, $dis$}}{
    $local\_die\_list$ = $expert\_die\_map$[$expert\_id$]\;
    $remote\_die\_list$ = \FindNearDies{$local\_die\_list$, $dis$}\;
    \Return $local\_die\_list + remote\_die\_list$\;
}

\end{algorithm}

\normalsize

\begin{figure}[t]
    \centering
    \includegraphics[width=0.49\textwidth]{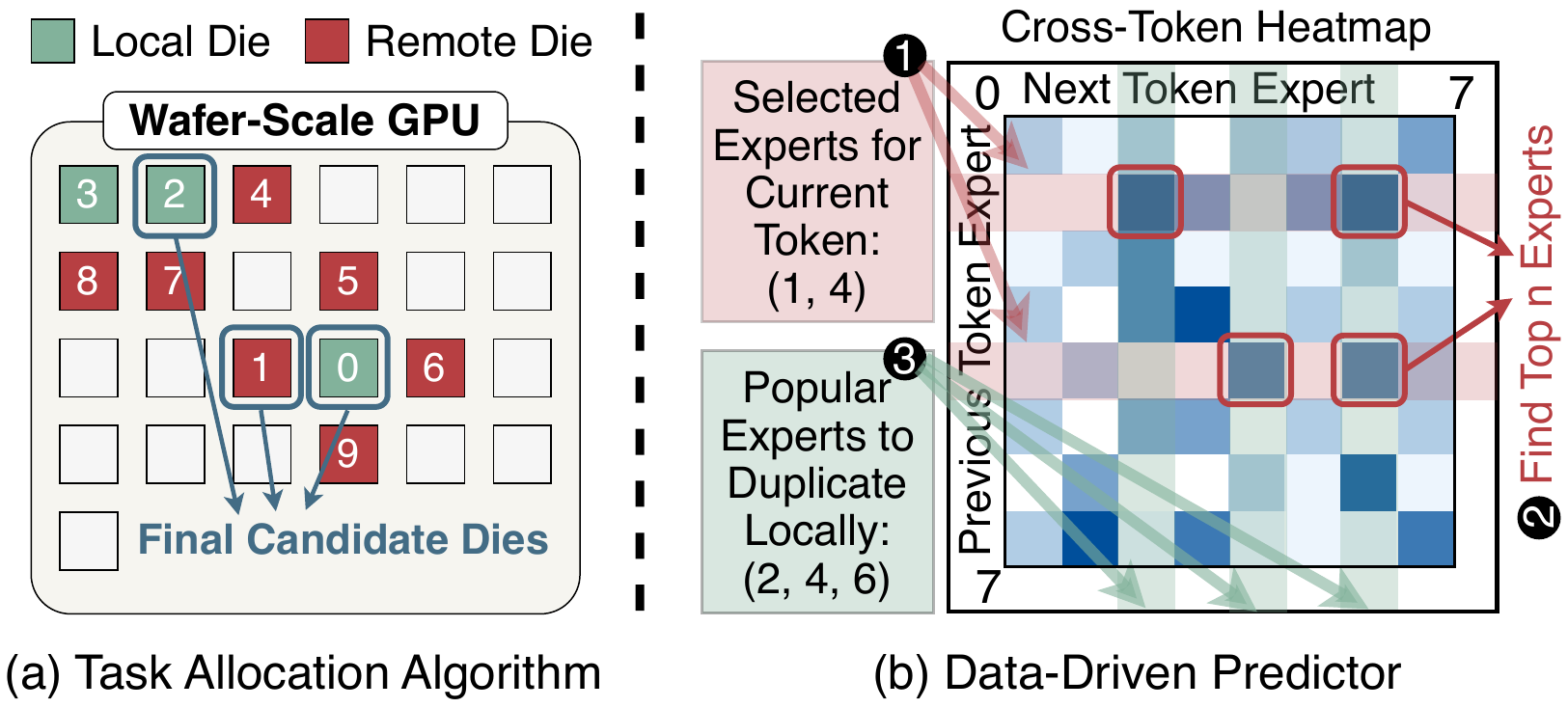}
    \caption{Proposed task allocation algorithm and data-driven predictor.}
    \label{fig:5_alg}
\end{figure}

The algorithm iterates through all experts to generate allocation plans. 
For each expert, it creates a candidate die list including dies storing expert weights and their adjacent dies (shown as green and red in \autoref{fig:5_alg}(a)). 
We sort candidates by workload and limit the list to $max_split_num$ dies, determined by the expert's request count (line 3-5). 
Requests are distributed to candidate dies in blocks of size 50 to balance efficiency and accuracy (line 6-11). 
For each block, the algorithm selects the optimal die using our cost model, which considers DRAM access, computation, and die-to-die communication. 
Finally, we merge blocks allocated to the same die to generate the final allocation plan (line 12).

\noindent\underline{Data-Driven Predictor: }
Our predictor algorithm, implemented by the global CP, uses current MoE kernel expert selection information to predict popular experts for the next token on each die. This prediction result is transferred to local CPs and configured in each die's PDU to guide hardware-managed local HBM.
As shown by the red shadow in ~\autoref{fig:5_alg}(b), we first identify corresponding rows from the heatmap based on current expert selection ({\small\circlednumber{black}{1}}), then select the top $n$ experts from each row ({\small\circlednumber{black}{2}}) and identify corresponding experts for the next token as prediction results, denoted by the green shadow ({\small\circlednumber{black}{3}}). 
In this example, a die computes experts 1 and 4 during the current MoE kernel and we predict experts 2, 4, and 6 will be used next. Since this die only reads experts 1 and 4 currently, we duplicate only expert 4 in its local DRAM.


\section{Evaluation}




\begin{figure*}[t!]
    \centering
    \includegraphics[width=0.99\textwidth]{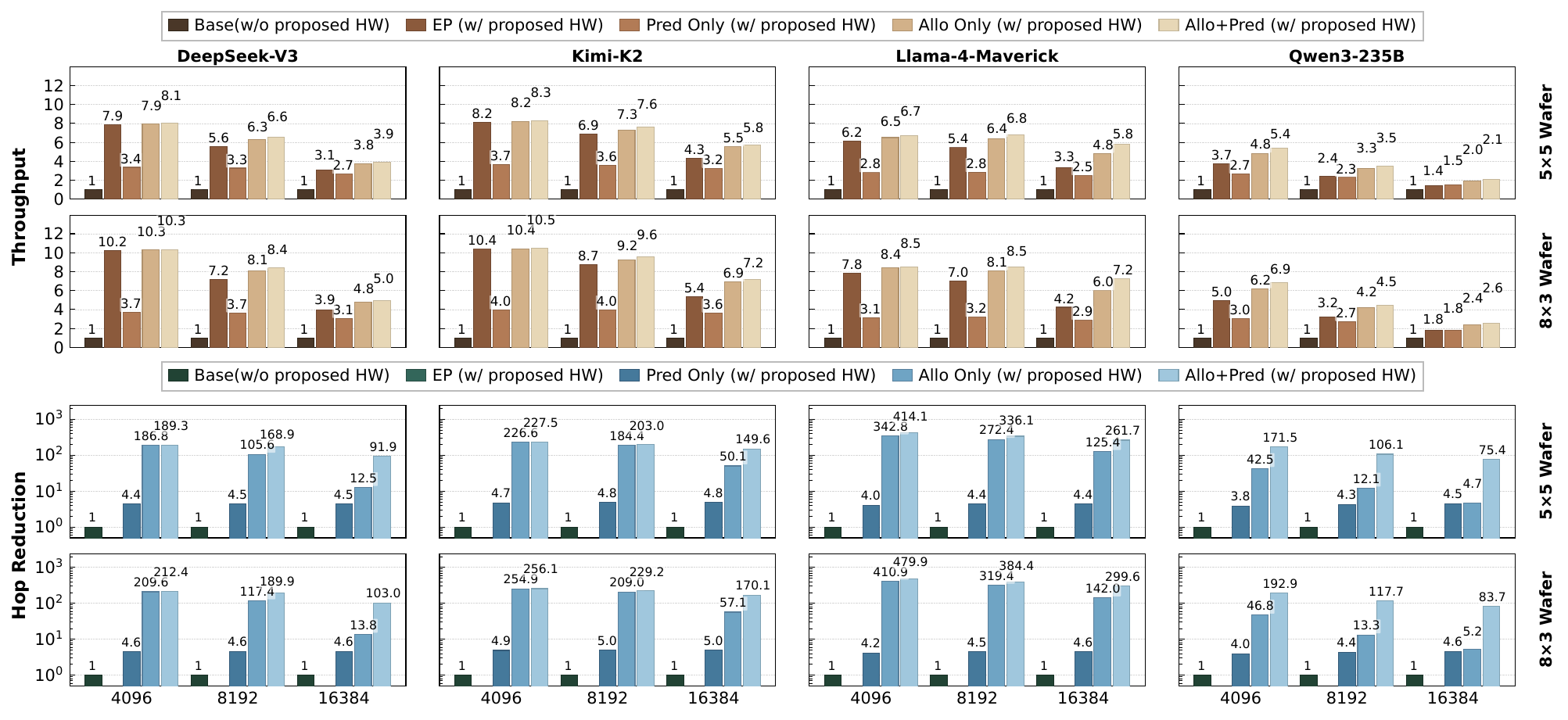}
    \caption{Throughput of MoE layers (Top) and hop number reduction ratio (Bottom). All figures are scaled to baseline.}
    \label{fig:exp_tp_hop}
\end{figure*}

\subsection{Experiment Setup}
\textbf{Methodology: }
We conduct experiments using event-driven simulation on a validated simulator. Expert selection traces are collected by deploying SGLang~\cite{zheng2024sglang} on an 8×H100 DGX server and an 8×H200 AWS instance. 


We developed a custom multi-chiplet GPU simulator in Python, as existing tools are inadequate for our needs. Cycle-accurate simulators such as Gem5~\cite{gem5}, gpgpusim~\cite{gpgpusim}, and mgpusim~\cite{mgpusim} accurately model single GPUs but are prohibitively slow for large-scale systems with 20+ dies and batch sizes exceeding 15,000. 
Event-driven simulators such as ASTRA-sim~\cite{astra} support multi-GPU systems but lack detailed microarchitecture modeling and do not support the single-GPU-like programming model we adopt. 
Our simulator models all key multi-chiplet GPU components, including LLC, HBM, compute units, and D2D links across all dies, with a central resource manager that captures contention and congestion. 
We validated the simulator against real measurements from an 8$\times$H100 DGX server, as detailed in ~\autoref{subsec: exp_validation_of_simulator}.


\begin{table}[t!]
\centering
\caption{Hardware Configurations}
\label{table:5_hardware_config}
\resizebox{0.48\textwidth}{!}{
\begin{tabular}{@{}ccccccc@{}}
\toprule
              & X-die                   & Y-die                   & \begin{tabular}[c]{@{}c@{}}DRAM \\      BW\end{tabular}                   & \begin{tabular}[c]{@{}c@{}}D2D \\      BW\end{tabular}                  & DRAM                    & \begin{tabular}[c]{@{}c@{}}Cmpt Power\\      per Die (FP16)\end{tabular}                  \\ \midrule
Dojo          & 5                       & 5                       & 3.35 TB/s                                                                    & 1.7 TB/s                                                                & 80GB                    & 989 TFLOPS                                                                              \\
TSMC-SoW      & 3                       & 8                       & 3.35 TB/s                                                                    & 1.7 TB/s                                                                & 80GB                    & 989 TFLOPS                                                                              \\
Dojo-Enhanced & 5                       & 5                       & 8 TB/s                                                                    & 2 TB/s                                                                  & 180GB                  & 4500 TFLOPS                                                                               \\ \midrule
Other Params  & \multicolumn{6}{c}{\begin{tabular}[c]{@{}c@{}}LLC hit latency:   100ns, LLC miss penalty: 110ns, \\      LLC write latency: 30ns, LLC size: 64 MB \\      D2D link latency: 200ns, Routing Alg: XY routing,\\      Command and address size for each remote request: 16B\\      Loca HBM access latency: 300 ns\end{tabular}} \\ \bottomrule
\end{tabular}
}
\end{table}

\color{black}

\textbf{Metric: } We measure the throughput of MoE layers during the decode stage as modern LLM serving systems show a trend toward fine-granularity disaggregation. Traditional LLMs benefit from separating prefill and decode stages across different machines, as demonstrated by DistServe~\cite{zhong2024distserve} and subsequent works~\cite{loongserve, qin2025mooncake}. For MoE models, this disaggregation extends further. MegaScale-Infer~\cite{zhu2025megascale-infer} separates attention and MoE operations onto different machines for optimal batch sizes. Following this trend, we focus on optimizing MoE operations during the decode stage.

\textbf{Hardware Configuration: }
We evaluate two multi-chiplet topologies: Tesla Dojo~\cite{dojo1,dojo2} and the TSMC SoW roadmap~\cite{tsmc-arch}. As summarized in \autoref{table:5_hardware_config}, Dojo uses a 5$\times$5 2D mesh, while TSMC SoW adopts an 8$\times$3 2D mesh. These choices reflect a deployed system (Dojo) and near-future industry support (TSMC SoW).

For both the Dojo and TSMC SoW configurations, each chiplet is H100-like, providing 1{,}000 TFLOPS FP16 compute, 80GB HBM, 3.35TB/s local HBM bandwidth, and 1.7~TB/s inter-die bandwidth to adjacent chiplets. 
We also include an extended experiment in ~\autoref{subsec: exp_cpu_overhead} with a Dojo-Enhanced configuration, where each die is B300-like to reflect an anticipated hardware performance trend in the future. We reserve 10\% of DRAM for system and hardware management.

\textbf{Baseline Configurations:}
We compare our approach against the simple strategy currently used by GPU. 

The \textbf{Base} configuration adopts an EP-like data placement and assigns an equal number of experts to each die. However, the entire wafer operates as a single large GPU: each die handles the same amount of expert computation without considering expert placement.

\textbf{EP} assigns each expert's computation to the die where it
resides, as also adopted by MoEntwine~\cite{tang2025moentwine}. This
eliminates all D2D communication but can cause severe workload imbalance.
Note that even under EP, our Global CP and Local CP architecture remains
necessary, as expert placement information is still required.

We implement three variants: \textbf{Allo Only} uses solely our task allocation strategy; \textbf{Pred Only} includes only the data-driven predictor; and \textbf{Allo+Pred} combines both techniques. These configurations evaluate the individual and combined effects of our proposed methods.

\textbf{Models and Workloads: }
We conduct evaluations with real traces collected from Qwen3 and Deepseek V3. The traces are gathered from diverse datasets, including MMLU~\cite{mmlu}, MMLU Pro~\cite{mmlu_pro}, ChineseSimpleQA~\cite{he2024chinese}, and LiveCodeBench~\cite{jain2024livecodebench}, comprising over 24,000 requests per model.
Each test batch is filled by sequentially adding requests in the order of
MMLU, MMLU-Pro (CH), ChineseSimpleQA, and LiveCodeBench until the target
batch size is reached.

\subsection{Validation of Simulator}
\label{subsec: exp_validation_of_simulator}
We validate our simulator using real measurements from an 8×H100 DGX server. We evaluate both single-GPU execution and two-GPU peer-to-peer (P2P) communication. 

For single-GPU execution, we benchmark one expert in a MoE layer,
which consists of three GEMM operations, across varying batch sizes for both DeepSeek and Qwen.

\begin{figure}[t!]
    \centering
    \includegraphics[width=0.48\textwidth]{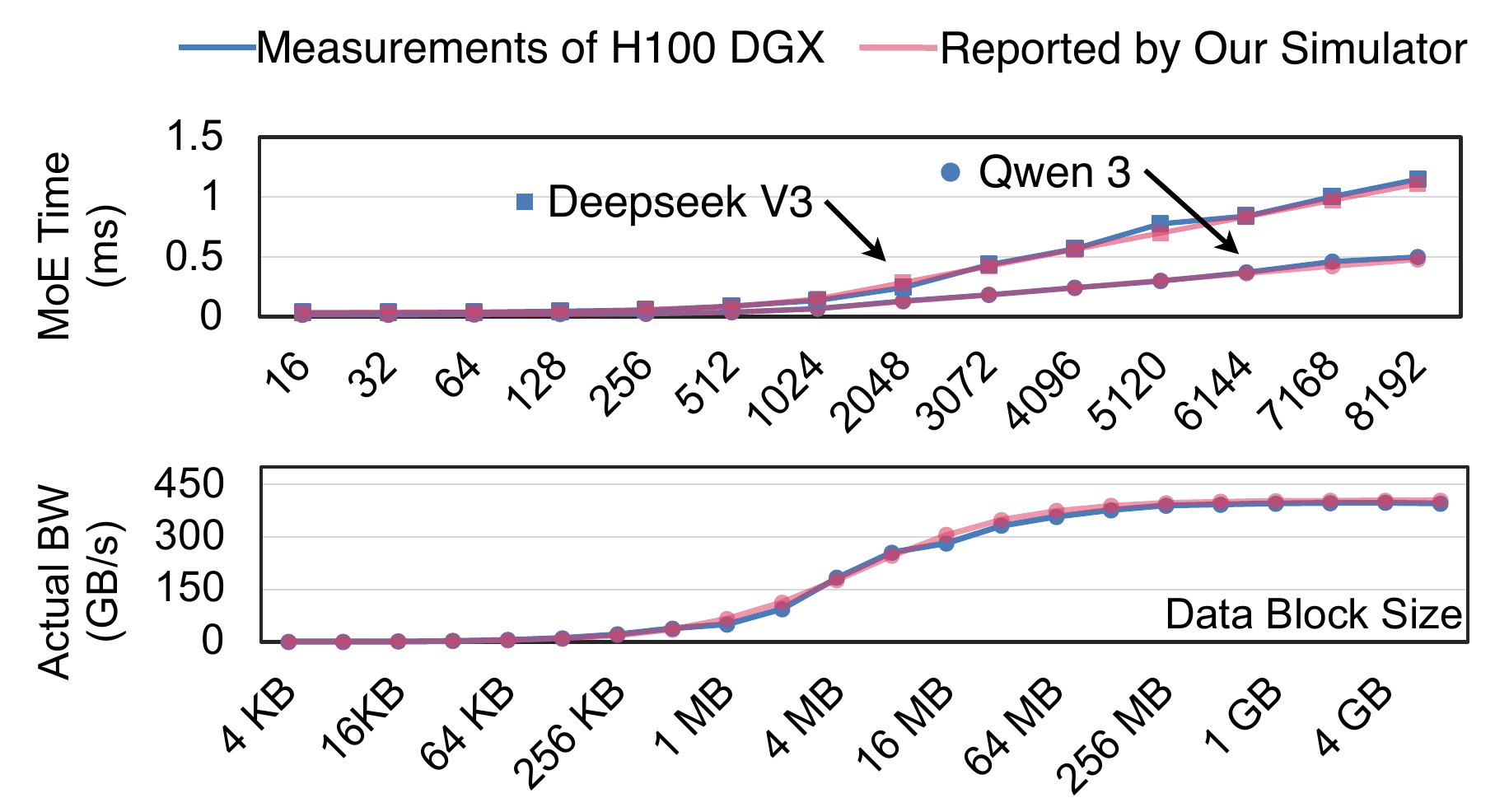}
    \caption{Simulator validation with real data generated from 8xH100 DGX, including both MoE Layer (Top) and P2P data transfer (Bottom) test cases.}
    \label{fig:exp_verify}
\end{figure}

For P2P communication, we measure data migration between two GPUs over payload sizes ranging from 4 KB to 4 GB. To ensure simulation fidelity, we calibrate key parameters to fit the measured data. As shown in ~\autoref{fig:exp_verify}, the simulator’s error remains within 5\% for all test cases.

\subsection{Throughput}
We evaluate MoE decode stage throughput in ~\autoref{fig:exp_tp_hop}, with results normalized to the baseline configuration.

\textbf{Comparison across models: }
Our Allo+Pred strategy achieves 7.0$\times$, 8.2$\times$, 7.3$\times$, and 4.1$\times$ throughput improvement on Deepseek, Kimi, Llama, and Qwen, respectively. Deepseek and Kimi benefit more due to their larger expert count (256 vs.\ 128) and more complex selection patterns.

%
\textbf{Comparison across chiplet architectures: }
Our strategy shows 6.0$\times$ improvement on Dojo and 7.5$\times$ on TSMC, despite similar die counts (25 vs.\ 24). TSMC's rectangular layout places dies farther apart, introducing more inter-unit communication without strategic task allocation, hence the larger gain under our strategy.

\textbf{Comparison with EP:}
At small batch sizes such as 4096, our strategy and EP perform similarly: few tokens per expert make execution memory-bound, so splitting one expert across multiple dies offers no benefit, and our strategy degenerates to EP. The advantage emerges at larger batches, achieving 1.44$\times$ speedup over EP at batch size 16{,}384.


\subsection{Hop Reduction}
We report hop counts in ~\autoref{fig:exp_tp_hop} to show the reduction in inter-unit communication. Hop count is the sum of Manhattan distances for all cross-unit communications. Higher hop counts indicate frequent cross-die data movement. We normalize results to baseline and report hop reduction ratios, where a ratio of 10 means the hop count is reduced to 1/10.

\textbf{Pred Only} reduces hop counts by 4.5$\times$, aligning with performance improvement of 3.0$\times$. This indicates cross-unit communication is the primary bottleneck in baseline, and reducing hop counts proportionally improves performance.

\textbf{Allo Only} reduces hop counts by 142$\times$, exceeding the performance improvement of 6.3$\times$. This shows that with our allocation algorithm, inter-unit communication is no longer the sole bottleneck. While reducing hop counts still improves performance, the improvement is not proportional.

\textbf{Allo+Pred} reduces hop counts by over 213$\times$ compared to baseline. However, performance improvement is only 6.63$\times$ over baseline, with just 1.1$\times$ average improvement over Allo Only. This demonstrates that hop count is no longer a performance bottleneck. With the help of our task allocation algorithm, most tasks are distributed to local dies holding related experts, with only extremely popular experts requiring remote allocation. This leads to minimal D2D traffic and shifts the bottleneck to workload distribution.

\subsection{DRAM Access Breakdown}
\begin{figure}[tbp]
    \centering
    \begin{minipage}{0.56\linewidth}
        \centering
        \includegraphics[width=1.0\textwidth]{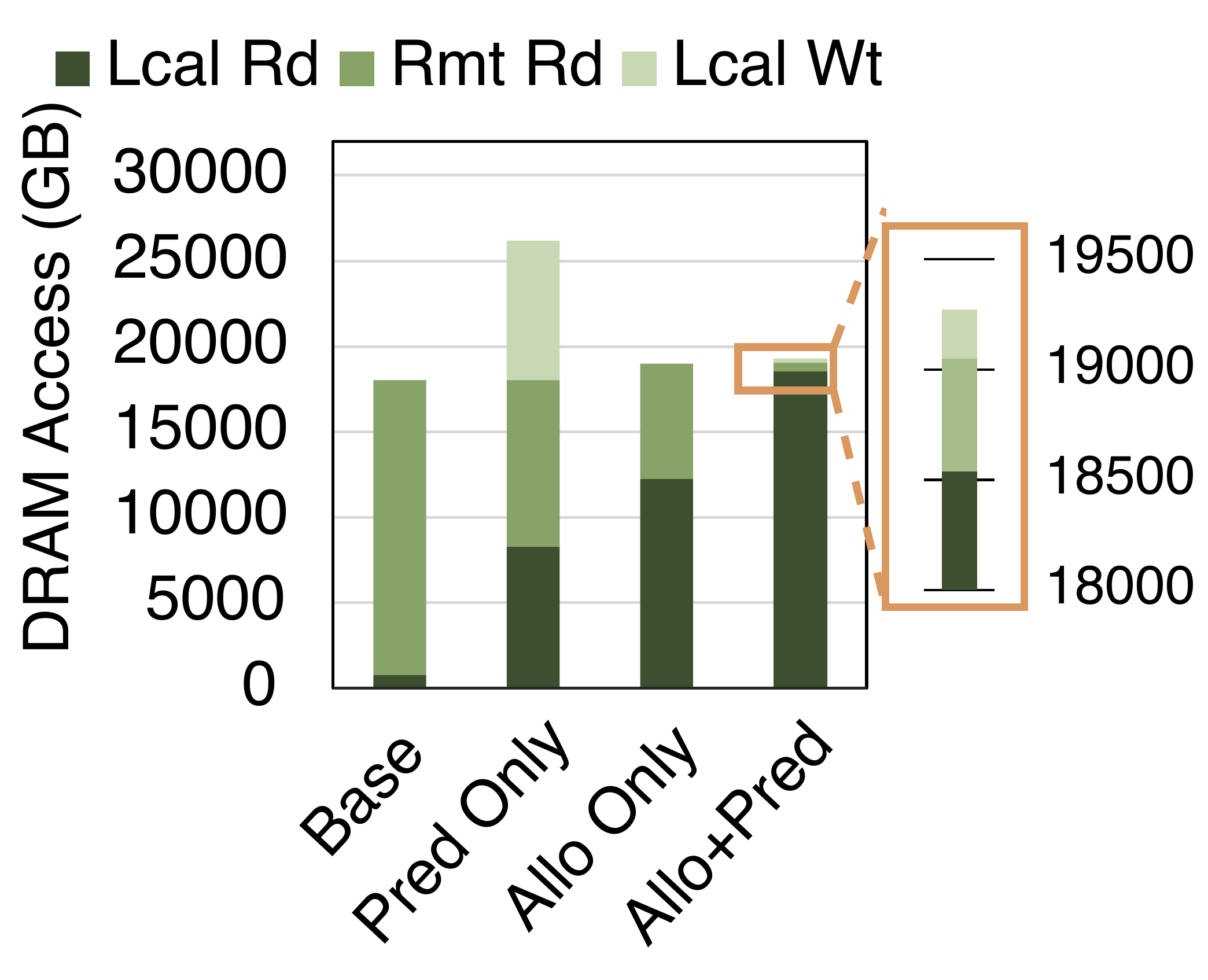}
        \caption{DRAM access breakdown for Qwen3 on TSMC-SoW Configuration with batch size 4096.}
        \label{fig:exp-breakdown}
    \end{minipage}
    \hfill 
    \begin{minipage}{0.405\linewidth} 
        \centering
        \includegraphics[width=1.0\textwidth]{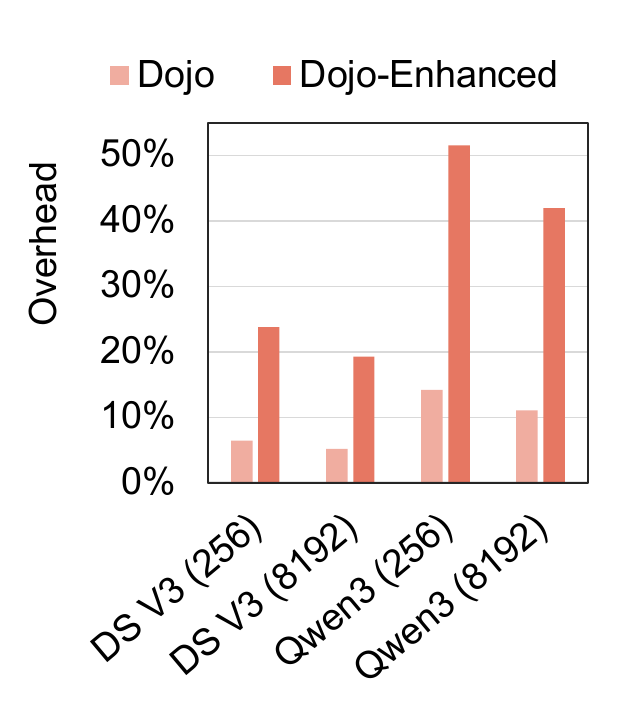}
        \caption{Host CPU implementation overhead under varying models and batch sizes.}
        \label{fig:exp_cpu_overhead}
    \end{minipage}
\end{figure}

We provide a breakdown of DRAM access patterns in ~\autoref{fig:exp-breakdown} to show how our strategies reduce inter-unit communication. 
We categorize DRAM access into three types: reads from local dies, reads from remote dies, and writes to local dies, where writes to local dies only occur when we duplicate a remote expert locally. 
Most reads in the baseline are from remote dies, resulting in high inter-unit traffic and poor performance. With our strategies (Pred Only, Allo Only, and Allo+Pred), most remote DRAM reads are converted to local DRAM reads, significantly reducing traffic. Compared with Pred Only, Allo+Pred achieves fewer remote reads by allocating most tasks to local dies, with only extremely popular experts requiring computation across multiple dies. Compared with Allo Only, Allo+Pred further reduces remote reads by caching popular experts in local HBM.


\subsection{Comparison with Host CPU-Based Implementation}
\label{subsec: exp_cpu_overhead}
Our task allocation algorithm runs on a new GPU command processor, but it could, in principle, be executed on the host CPU with a higher overhead.
As shown in ~\autoref{fig:exp_cpu_overhead}, we evaluate both the Dojo and Dojo-Enhanced configurations. In Dojo, the overhead of host-CPU allocation is 5.2\%–6.4\% for DeepSeek V3 and 11.1\%–14.2\% for Qwen3. In Dojo-Enhanced, the overhead rises to 19.3\%–23.8\% for DeepSeek V3 and 42.0\%–51.6\% for Qwen3.

\textbf{DeepSeek vs Qwen:} Qwen3 incurs higher overhead than DeepSeek V3 due to CPU–GPU data transfers over PCIe, which occur once per MoE layer. The CPU needs the Expert Distribution Table from the GPU to run the allocator, and the allocation results must be sent back to the GPU before kernel execution. Qwen3 has (i) more MoE layers (94 vs. 58), increasing transfer frequency, and (ii) smaller per-layer compute, which amplifies the relative cost of transfers.

\textbf{Dojo vs Dojo-Enhanced:} Dojo-Enhanced shows over 3.7× higher overhead than Dojo because its GPU dies are significantly faster, making fixed PCIe transfer costs dominate more. As GPU performance outpaces interconnect bandwidth, implementing the allocator in the GPU command processor becomes increasingly necessary to sustain performance.

\subsection{Area and Power Overhead}


We estimate the area and power overhead of all added modules in~\autoref{table:overhead}. Our design supports up to 100 layers with 512 experts per layer, well beyond SOTA MoE model (Kimi-K2: 61 layers, 384 experts). The full heatmap (50\,MB) is stored in Global CP DRAM, with a 0.5\,MB on-chip cache buffering one layer at a time. The Prediction Table is implemented in registers due to its small size; all other components use SRAM. Registers are synthesized with Yosys~\cite{wolf2013yosys} and SRAM is modeled with CACTI~\cite{balasubramonian2017cacti}, both scaled to 5\,nm to match the H100 process node. Global and Local CP area estimates are derived from ARM core data. As shown in~\autoref{table:overhead}, total area and power overhead is less than 0.04\%.

\begin{table}[h]

\centering
\caption{Area and Power Overhead.}
\label{table:overhead}

\resizebox{0.99\linewidth}{!}{%
\begin{tabular}{@{}cccccc@{}}
\toprule
\multicolumn{1}{l}{}      & Capacity & \begin{tabular}[c]{@{}c@{}}Bit \\ Width\end{tabular} & \begin{tabular}[c]{@{}c@{}}Num\\ Per Wafer\end{tabular} & \begin{tabular}[c]{@{}c@{}}Tot Area\\ (mm2)\end{tabular} & \begin{tabular}[c]{@{}c@{}}Tot Power\\ (mW)\end{tabular} \\ \midrule
Prediction Table          & 128 B    & 16 bit                                               & 25                                                      & 0.0020                                                   & 55.75                                                    \\
Address Translation Unit  & 4.25 KB  & 68 bit                                               & 25                                                      & 0.0048                                                   & 334.25                                                   \\
Local CP (A72)~\cite{a72}            & N/A      & N/A                                                  & 25                                                      & $\sim$7.5000                                                   & $\sim$7000                                                    \\
Expert Distribution Table & 4.5 KB   & 72 bit                                               & 1                                                       & 0.0002                                                   & 13.94                                                    \\
Heatmap Cache             & 0.5 MB   & 512 bit                                              & 1                                                       & 0.0278                                                   & 184.67                                                   \\
Global CP (A76)~\cite{a76}           & N/A      & N/A                                                  & 1                                                       & $\sim$1.1000                                                   & $\sim$1000                                                     \\ \midrule
Total                     &          &                                                      &                                                         & 6.13                                                     & 8588.61                                                   \\
Overhead (25-die wafer)                  &          &                                                      &                                                         & $\sim$0.04\%                                             & $\sim$ 0.04\%                                        \\ \bottomrule
\end{tabular}
}
\end{table}

\begin{figure}[t!]
    \centering
    \includegraphics[width=0.47\textwidth]{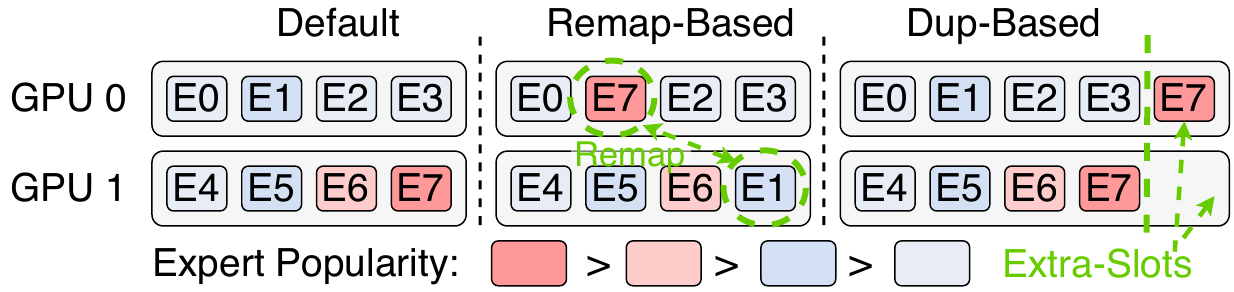}
    \caption{Demonstration of expert placement strategies.}
    \label{fig:5_expert_placement_demo}
\end{figure}

\section{Case Study 2: Prefill-Guided Decode Expert Placement on Real GPU Clusters}
\label{sec: case study 2}
\subsection{Introduction}
Workload imbalance is one of the biggest challenges in large-scale MoE serving (200+ GPUs). EPLB~\cite{deepseek_eplb_2026} addresses this by dynamically adjusting expert placement, but it is triggered every 3000+ steps and relies on periodically collected profiling data~\cite{vllm_eplb_2025}. A natural question then arises: how to set expert placement for the initial $\sim$1000 decode tokens when no profiling data are yet available? This is especially pressing for short-output requests, for which EPLB never collects enough data to be effective. Inspired by \hyperref[box:insight1]{\textcolor{magenta}{\underline{Insight 1}}}, which reveals temporal correlation between prefill and decode stages, we propose leveraging prefill-stage expert selection information to guide expert placement for initial decode steps.

As shown in~\autoref{fig:5_expert_placement_demo}, we design two placement algorithms (details in~\autoref{alg:prefill-placement}). The \emph{Remap-based} algorithm keeps the number of experts per GPU unchanged and reassigns experts across GPUs for a more balanced workload: it sorts experts by decreasing roofline cost and greedily assigns each to the least-loaded GPU, subject to a uniform capacity of $E/G$ experts per GPU. The \emph{Duplication-based} algorithm reserves extra expert slots on each GPU and uses prefill traces to duplicate hot experts, thereby avoiding congestion: starting from the default contiguous layout (e.g., experts 0--15 on GPU,0, 16--31 on GPU,1, etc.), it greedily adds up to $R$ extra replicas per GPU, selecting at each step the (expert, GPU) pair that maximally reduces the bottleneck load $\max_g \mathit{load}_g$; tokens of a replicated expert are evenly split among all its copies. Both algorithms use a roofline-based cost model to estimate per-GPU load.

\SetKwFunction{RemapPlacement}{remap\_based\_placement}
\SetKwFunction{DupPlacement}{dup\_based\_placement}
\begin{algorithm}[t]

\footnotesize
\LinesNumbered
\caption{Prefill-Guided Expert Placement}
\label{alg:prefill-placement}

\KwIn{Prefill traces $\mathcal{D}$, GPU count $G$, extra slots per GPU $R$}
\KwOut{Per-layer expert-to-GPU assignment $\{\mathcal{S}_g\}_{g=1}^{G}$}

\textbf{Notation:} $E$: total experts; $f_{l,e}$: freq of expert $e$ at layer $l$; $L_g$: load of GPU $g$; $r_g$: remaining slots on GPU $g$; $\delta_{e,g}$: $\max_{g'}L_{g'}$ change after copying expert $e$ to GPU $g$

\BlankLine

\SetKwProg{Fn}{Function}{:}{}
\Fn{\RemapPlacement{$\mathcal{D}, G$}}{
    \For{each layer $l$}{
        Compute $f_{l,e}$ from $\mathcal{D}$; sort exps by decreasing $\mathrm{Cost}(f_{l,e})$\;
        $L_g \gets 0$ for all $g$\;
        \For{each expert $e$ in sorted order}{
            Assign $e$ to least-loaded GPU $g^*$ s.t.\ $|\mathcal{S}_{g^*}|<E/G$; $L_{g^*} \mathrel{+}= \mathrm{Cost}(f_{l,e})$\;
        }
    }
    \Return $\{\mathcal{S}_g\}$ for each layer\;
}

\BlankLine

\Fn{\DupPlacement{$\mathcal{D}, G, R$}}{
    \For{each layer $l$}{
        Compute $f_{l,e}$ from $\mathcal{D}$; generate default placement $\mathcal{S}_g$\;
        $r_g \gets R$ for all $g$\;
        $L_g \gets \sum_{e \in \mathcal{S}_g} \mathrm{Cost}(f_{l,e})$ for all $g$\;
        \For{$i \gets 1$ \KwTo $R \cdot G$}{
            $(e^*,g^*) \gets \arg\min_{e,g:\; r_g>0,\; g\notin\mathrm{hosts}(e)} \delta_{e,g}$\;
            Assign $e^*$ to $\mathcal{S}_{g^*}$; $r_{g^*} \gets r_{g^*}-1$\;
            update affected $L_g$\;
        }
    }
    \Return $\{\mathcal{S}_g\}$ for each layer\;
}

\end{algorithm}

\subsection{Methodology}
We deploy Qwen3-235B with SGLang on 8$\times$H100 GPUs with NVLink. We build a distributed profiler by inserting \texttt{cuda.Event} timers into SGLang to measure individual operations (attention, top-$k$, all-to-all, and MoE) on each GPU independently. 
We manipulate expert placement through SGLang's \texttt{init\_expert\_location} interface and use \texttt{DeepEP} as the MoE backend. 
The \texttt{ep\_dispatch\_algorithm} is set to ``dynamic'' so that tokens are evenly distributed across replicas of a duplicated expert.


\textbf{Metric.}
We report MoE computation time, i.e., all three expert linear layers, excluding attention, all-to-all, and top-$k$.

\textbf{Model and Benchmark.}
We evaluate on Qwen3-235B (94 MoE layers, 128 experts per layer, 8 selected). We use MMLU and Global-MMLU datasets, following the original ordering. Batch sizes range from 64 to 16{,}384.

\textbf{Baselines.}
\texttt{Default} is the standard contiguous placement used by Qwen and SGLang (experts 0--15 on GPU-0, 16--31 on GPU-1, etc.). \texttt{Best} and \texttt{Worst} are the theoretically optimal and worst placements generated with oracle decode-stage selections (not available in practice). \texttt{Remap} and \texttt{Dup} are our two prefill-guided strategies. For Dup, we use one extra slot per GPU, yielding 128+8=136 experts per layer.

\subsection{Results}

\begin{figure}[t!]
    \centering
    \includegraphics[width=0.42\textwidth]{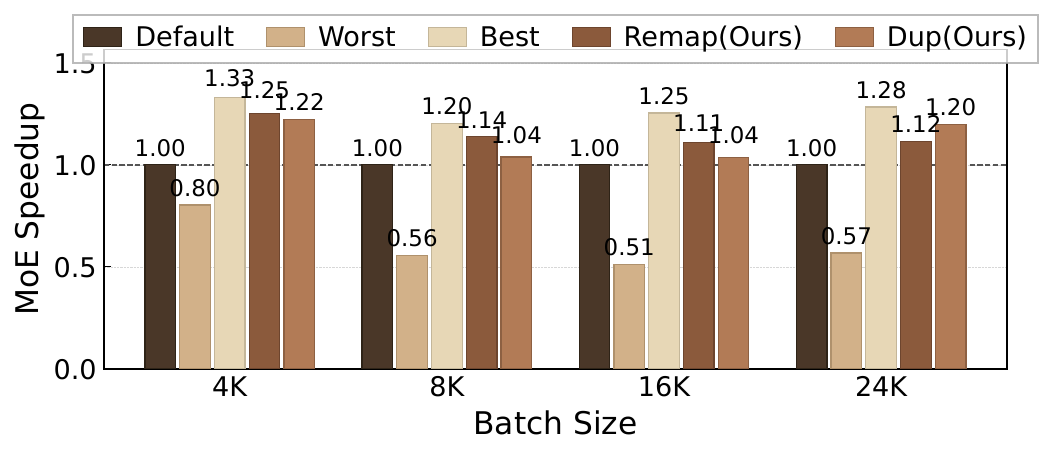}
    \caption{Performance of our prefill-aware expert placement.}
    \label{fig:exp_remap_moe_speedup}
\end{figure}

As shown in~\autoref{fig:exp_remap_moe_speedup}, \texttt{Remap} and \texttt{Dup} achieve speedups of 15.5\% and 12.5\% over \texttt{Default}, respectively, and deliver over 2$\times$ speedup compared with \texttt{Worst}. Both remain within 10\% of \texttt{Best}, which exploits oracle decode-stage information unavailable in practice, which demonstrates the effectiveness of our approach. Since the two algorithms perform comparably, one can choose between them to fit different memory and system constraints.

We note that our 8-GPU EP scale inherently limits the achievable improvement: with EP8, each GPU holds 16 experts per layer, so every GPU likely contains a mix of hot and cold experts, naturally yielding a relatively balanced workload even under the \texttt{Default} layout (the max/min execution-time ratio is only about 1.3$\times$). We expect greater speedups at larger EP scales where load imbalance is more pronounced.

\section{Discussion}


Both the wafer-scale GPU architecture and the prefill-guided expert placement strategy serve as case studies demonstrating the practical applicability of our profiling insights, which constitute the paper's primary contribution. Specifically, the wafer-scale GPU design follows \hyperref[box:insight3]{\textcolor{magenta}{\underline{Insight 3}}} for task allocation and leverages temporal relation insights (\hyperref[box:insight1]{\textcolor{magenta}{\underline{Insight 1}}} and part of \hyperref[box:insight2]{\textcolor{magenta}{\underline{Insight 2}}}) to build a data-driven predictor. The prefill-guided placement strategy utilizes \hyperref[box:insight1]{\textcolor{magenta}{\underline{Insight 1}}} to guide decode-stage expert placement using information collected during prefill.

Importantly, our insights extend far beyond these two case studies and can benefit a wide range of MoE serving systems, including multi-GPU clusters (Multi-Node DGX~\cite{zhu2025megascale-infer, zhang2025comet, chen2022tamoe} and NVL72~\cite{li2026dwdp}), CXL-/CPU-based memory disaggregation~\cite{moelightning,fang2025klotski}, flash-based multi-tier systems~\cite{cambricon-llm, sheng2023flexgen}, PIM architectures~\cite{yun2024duplex, pan2025stratum, yu2026ammamultichipletmemorycentricarchitecture}, and other emerging platforms.




\section{Related Work}

\textbf{MoE model behavior studies: }
Several MoE model tech reports~\cite{muennighoff2024olmoe, jiang2024mixtral, dai2024deepseekmoe, sun2024hunyuan} provide the MoE routing patterns as part of their evaluation.
For example, the Mixtral report~\cite{jiang2024mixtral} shows the temporal locality of expert selection by reporting the percentage of repetitive assignment.
The OLMoE report~\cite{muennighoff2024olmoe} shows the co-activation pattern and domain specialization among the experts.
A blog post from SGLang~\cite{sglang-deepseek-blogpost} shows expert distribution statistics for the DeepSeekV3 model and the inherent imbalance in expert selection and similarity between prefill and decode.
None of these studies provides a comprehensive profiling across multiple large-scale ($>$200B) MoE models, nor do they present a data-movement-centric methodology to highlight the opportunities like this paper.
\\
\textbf{Data movement optimization for MoE LLM inference}
Various prior works~\cite {hwang2024pre, li2023accelerating, gupta2024lynx, kamahori2025fiddler, eliseev2023fast, du2024sida, li2024optimizing,  skliar2024mixture, moelightning, yun2024duplex, zhang2025comet, hwang2023tutel} have focused on improving the efficiency of MoE LLM inference by reducing data movement.
For example, Lina and SmartMoE~\cite{li2023accelerating, zhai2023smartmoe} exploit expert selection skewness to dynamically schedule resources during inference and balance traffic across GPUs.
LYNX~\cite{gupta2024lynx} dynamically reduces active experts while preserving model accuracy.
Pre-gate MoE~\cite{hwang2024pre} uses a pre-gating function to alleviate the dynamic nature of expert selection.
Sida~\cite{du2024sida} builds an offline hash function to predict expert usage and reduce data movement between CPU and GPU.
MoE-Lightning~\cite{moelightning} leverages a CPU-GPU pipeline and paged weights to improve resource utilization.
Eliseev and Mazur~\cite{eliseev2023fast} exploit expert locality and leverage LRU caching to manage GPU and CPU memory.
This work targets data movement reduction in MoE LLM inference. Our cross-model profiling reveals optimization principles that apply broadly to current and future systems regardless of scale.
\\
\textbf{Wafer-Scale and Chiplet Architectures}
As single-chip scaling slows, wafer-scale and chiplet packaging offer promising paths for improved compute efficiency. Prior work targets either interconnect design~\cite{rashidi2025fred, yang2025pd, yu2025cramming, li2024lucie, chen2024waferscale, feng2023heterogeneous, feng2023scalable} or data placement for specific algorithms~\cite{he2025waferlm, tan2021nn} and applications~\cite{shao2019simba, xu2025wsc, odema2024scar}. In contrast, we are the first to study MoE LLM serving on wafer-scale GPUs and propose data-movement-centric HW/SW co-design optimizations.

\section{Conclusion}

Unlike prior MoE serving studies that take system-centric approaches with deployment-specific strategies, we study MoE serving from a model-focused perspective. We conduct comprehensive data-movement-centric profiling of state-of-the-art MoE models (200B–1000B) to extract system-agnostic insights, revealing structured patterns underlying seemingly random data movement and providing actionable guidance on system design. 
We validate these insights on both a future wafer-scale GPU architecture and existing multi-GPU systems, achieving significant performance improvements through minimal architectural modifications and lightweight software design, demonstrating their broad applicability.

\section*{Acknowledgment}
We thank all reviewers for their constructive feedback and insightful suggestions. This work is partially supported by Samsung Semiconductor.
\section*{Artifact Appendix}

\subsection*{A.1\quad Abstract}
This artifact packages the code, traces, scripts, and plotting utilities for reproducing the paper's main results across the two case studies.

\textbf{Case Study 1} is a CPU-runnable wafer-scale GPU simulator for MoE inference. It evaluates our expert allocation and prediction strategies across four large-scale MoE models and two chiplet topologies, reproducing Figure~\ref{fig:exp_tp_hop}.

\textbf{Case Study 2} contains the real-GPU expert placement experiments and reproduces Figure~\ref{fig:exp_remap_moe_speedup} on an 8$\times$H100 system. It requires a specialized GPU software stack. Both artifacts provide a \texttt{main\_ae.py} workflow for downloading traces, running experiments, and generating figures.

\subsection*{A.2\quad Artifact Check-List}
\noindent\textbf{Program:} Python~3.\\
\textbf{Run-time environment:} Linux with Python~$\geq$~3.10.\\
\textbf{Hardware:} Case Study 1 requires a CPU server with $\geq$64\,GB RAM. Case Study 2 requires an 8$\times$NVIDIA H100 80\,GB GPU server.\\
\textbf{Output:} Paper figures and CSV result files.\\
\textbf{Disk space:} About 80\,GB for one model and up to 300\,GB for all four models.\\
\textbf{Experiment time:} Case Study 1 takes 8--12 hours for one model or 18--36 hours for all models. Case Study 2 takes 12--16 hours.\\
\textbf{Publicly available:} Yes.\\
\textbf{Code license:} Apache-2.0.

\subsection*{A.3\quad Description}

\subsubsection*{A.3.1\quad How to Access}
\noindent\textbf{Case Study 1 (wafer-scale GPU simulator):} GitHub: \href{https://github.com/zhongkaiyu/waferscale_gpu_moe_sim}{\mbox{\texttt{waferscale\_gpu\_moe\_sim}}}; DOI: \href{https://doi.org/10.5281/zenodo.19617713}{\mbox{10.5281/zenodo.19617713}}.

\noindent\textbf{Case Study 2 (real-GPU expert placement):} GitHub: \href{https://github.com/zhongkaiyu/moe_exp_placement}{\mbox{\texttt{moe\_exp\_placement}}}; DOI: \href{https://doi.org/10.5281/zenodo.19617695}{\mbox{10.5281/zenodo.19617695}}.

Each repository contains a \texttt{README.md} with setup, execution, and troubleshooting instructions. The Zenodo archives provide persistent snapshots of the evaluated artifact versions.

\subsubsection*{A.3.2\quad Hardware Dependencies}
Case Study 1 runs on a CPU server with at least 64\,GB RAM and does not require a GPU. Case Study 2 requires an 8$\times$NVIDIA H100 80\,GB GPU server, CUDA~12.0 or newer, and about 300\,GB of disk space. Reviewers without GPU access can still evaluate the primary simulator artifact.

\subsubsection*{A.3.3\quad Software Dependencies}
Case Study 1 requires Python~$\geq$~3.10 plus \texttt{numpy}, \texttt{pandas}, and \texttt{matplotlib}; the scripts install them automatically. Case Study 2 additionally requires PyTorch, a modified SGLang fork, DeepEP, and DeepGEMM. The repository documents exact installation commands and environment settings.

\subsubsection*{A.3.4\quad Datasets}
Both artifacts use pre-recorded MoE expert-selection traces from MMLU. The traces are hosted on HuggingFace and downloaded automatically by the AE scripts.

\subsection*{A.4\quad Installation}
Installation instructions are provided in each repository's \texttt{README.md}. Case Study 1 is self-contained and intended as the default AE path. Case Study 2 includes GPU stack and communication library setup.

\subsection*{A.5\quad Experiment Workflow}
Both repositories provide a \texttt{main\_ae.py} entry point. The script downloads traces, runs experiments, collects CSV files, and regenerates the corresponding paper figure. Reviewers may also run individual model configurations using the repository \texttt{README.md} commands.

\subsection*{A.6\quad Evaluation and Expected Results}
Case Study 1 reproduces Figure~\ref{fig:exp_tp_hop}. The simulator is deterministic, so generated results should match the reported trends when using the same traces and configuration files.

Case Study 2 reproduces Figure~\ref{fig:exp_remap_moe_speedup}. Because it measures real GPU execution, small timing variations are expected from thermals, system load, NCCL non-determinism, and SGLang micro-batching. In our runs, variation is typically within $\pm$5\%. The high-level result is stable: prefill-aware placement improves MoE kernel performance by about 5--25\% over the default placement.

\subsection*{A.7\quad Methodology}
Submission, reviewing, and badging follow the \href{https://www.acm.org/publications/policies/artifact-review-and-badging-current}{ACM artifact review guidelines} and the \href{https://cTuning.org/ae}{cTuning AE guidelines}.

\bibliographystyle{IEEEtran}
\bibliography{main}

\end{document}